\pdfoutput=1
\documentclass{jfm}
\usepackage{graphicx}
\usepackage{subfigure,rotating,amsmath,float}
\usepackage{IEEEtrantools}
\usepackage{natbib}
\usepackage{color}
\ifCUPmtlplainloaded \else
  \checkfont{eurm10}
  \iffontfound
    \IfFileExists{upmath.sty}
      {\typeout{^^JFound AMS Euler Roman fonts on the system,
                   using the 'upmath' package.^^J}%
       \usepackage{upmath}}
      {\typeout{^^JFound AMS Euler Roman fonts on the system, but you
                   dont seem to have the}%
       \typeout{'upmath' package installed. JFM.cls can take advantage
                 of these fonts,^^Jif you use 'upmath' package.^^J}%
      }
  \else
  \fi
\fi

\ifCUPmtlplainloaded \else
  \checkfont{msam10}
  \iffontfound
    \IfFileExists{amssymb.sty}
      {\typeout{^^JFound AMS Symbol fonts on the system, using the
                'amssymb' package.^^J}%
       \usepackage{amssymb}%
         \let\leq=\leqslant
         
      }{}
  \fi
\fi

\ifCUPmtlplainloaded \else
  \IfFileExists{amsbsy.sty}
    {\typeout{^^JFound the 'amsbsy' package on the system, using it.^^J}%
     \usepackage{amsbsy}}
    {\providecommand\boldsymbol[1]{\mbox{\boldmath $##1$}}}
\fi

\newsavebox{\astrutbox}
\sbox{\astrutbox}{\rule[-5pt]{0pt}{20pt}}

\newcommand\bdel{\ensuremath{\boldsymbol{\delta}}}
\newcommand\bxm{\ensuremath{\mathbf{x}_{\mathrm{m}}}}

\newcommand\bxprm{\ensuremath{\mathbf{x}^{\prime}}}
\newcommand\bx{\ensuremath{\mathbf{x}}}
\newcommand\bxzero{\ensuremath{\mathbf{x}_{0}}}

\newcommand\bfor{\ensuremath{\mathbf{f}}}
\newcommand\bu{\ensuremath{\mathbf{u}}}
\newcommand\bul{\ensuremath{\mathbf{u}^{\mathrm{l}}}}
\newcommand\bug{\ensuremath{\mathbf{u}^{\mathrm{g}}}}
\newcommand\bubx{\ensuremath{\mathbf{u}\left(\bx\right)}}
\newcommand\alc{\ensuremath{\alpha_{\mathrm{cut}}}}

\newcommand\xmy{\ensuremath{\mathbf{x}^{\prime}-\mathbf{y}^{\prime}}}
\newcommand\bfe{\bfor^{\mathrm{e}}}
\newcommand\brho{\ensuremath{\boldsymbol{\rho}}}
\newcommand\brhol{\ensuremath{\boldsymbol{\rho}^{\mathrm{l}}}}
\newcommand\brhog{\ensuremath{\boldsymbol{\rho}^{\mathrm{g}}}}
\newcommand\lp{\ensuremath{\left(}}
\newcommand\rp{\ensuremath{\right)}}
\newcommand\bxpp{\ensuremath{ \lp \bx \rp}}

\newcommand\bxmpp{\ensuremath{ \lp \bxm \rp}}
\newcommand\NS{\ensuremath{N_{\mathrm{SH}}}}

\newcommand\bG{\ensuremath{ \mathbf{G}}}
\newcommand\np{N^{\mathrm{p}}}
\newcommand\rcut{R_{\mathrm{cut}}}
\newcommand\N{N_{\mathrm{d}}}
\newcommand\bQ{\mathbf{Q}}
\newcommand\bT{\mathbf{T}}

\def \tpmax {\tau^{P}_{\mathrm{max}}}
\def \tpmaxiso {\tau^{P_{\mathrm{ISO}}}_{\mathrm{max}}}
\def \gs {G_{s}}

\title[Deforming capsule through a corner]{The motion of a deforming capsule\\
through a corner}

\author[L. Zhu and L. Brandt]%
{Lailai Zhu$^{1,2}$%
  \thanks{Email address for correspondence: lailaizhu00@gmail.com} \ns
and Luca Brandt$^1$}

\affiliation{$^1$Swedish e-Science Research Centre and Linn\'e Flow Centre,\\ KTH Mechanics, S-100 44 Stockholm, Sweden
\\[\affilskip]
$^2$ Laboratory of Fluid Mechanics and Instabilities,\\
Station 9, EPFL, 1105 Lausanne, Switzerland}

\pubyear{2010}
\volume{650}
\pagerange{119--126}
\date{}

\begin{document}
\maketitle 

\begin{abstract}
A three-dimensional deformable capsule convected through a
square duct with a corner is studied via numerical simulations.
We develop an accelerated boundary integral implementation
adapted to general geometries and boundary conditions. 
A global spectral method is adopted to resolve the dynamics of the capsule membrane developing elastic tension 
according to the neo-Hookean constitutive
law and bending moments in an inertialess flow.
The simulations show that the trajectory of the capsule closely follows the underlying
streamlines independently of the capillary number. The membrane deformability, on the other hand, 
significantly 
influences the relative area variations, the advection velocity and the principal tensions observed during the capsule 
motion.
The evolution of the capsule velocity displays a loss of the time-reversal symmetry of Stokes flow 
due to the elasticity of the membrane. The velocity decreases while the capsule is approaching the corner
as the background flow does, reaches a minimum at the corner
and displays an overshoot past the corner due to the streamwise elongation induced by the flow acceleration in the downstream branch. This velocity 
overshoot increases with confinement while the maxima of the major principal tension increase linearly 
with the inverse of the duct width. Finally, the deformation and tension of the capsule are shown to decrease in a curved corner.

\end{abstract}


\section{Introduction}
Elastic micro-capsules are ubiquitous in nature, appearing in the form
of seeds, eggs, cells and similar. The elasticity of the cells plays an
important role for their proper biological functioning. As examples, red
blood cells (RBC) deform significantly in micro vessels to ease oxygen
transportation; leukocytes squeeze through small gaps into the
endothelial cell wall during inflammation~\citep{leu_deform} so as
tumor cells do in tumor metastasis~\citep{tumor_deform}. On the
other hand, artificial micro-capsules are commonly used in the food and
cosmetic industry for a controlled release of
ingredients~\citep{dbb11capreview} and synthetic nano-capsules promise a 
precise and targeted drug
delivery. The ability of biological and artificial capsules to
dynamically adapt, 
change their shapes and
withstand
stresses from the surrounding medium has thus attracted remarkable
attention from research groups in different fields.

In micro-fluidic applications, one of the most fundamental issues is the
behaviour of these tiny deformable structures when interacting with an external
applied flow. Early experimental studies discovered several
interesting features of RBCs: the well-known tank-treading and
tumbling motion in shear
flow~\citep{gold1972,fischer1978tank},
'parachute' shaped deformation~\citep{skalak1969parashu} and the 'zipper'
flow pattern~\citep{1980zipperflow} in the micro-capillaries. These
observations show that the capsule shape is not given \textit{a priori}
but determined by the dynamic balance of interfacial forces with
fluid stresses. 
Several analytical studies deal with
unbounded domains to model of tank-treading and
tumbling motions of an initially spherical capsule by asymptotic
analysis~\citep{dbb1980_asym,dbb1981time}; prove the existence of
'slipper' shaped cells in capillary flows~\citep{secomb1982_slipper};
predict of the vacillating-breathing behaviour of a vesicle \citep{misbah06_breath} and 
the swinging-tumbling transition of a capsule~\citep{vlahovska2011dynamics}.

Numerical simulations have been successfully used to solve
the associated nonlinear fluid-structure problem; examples are the deformation of
spherical~\citep{poz95shear,poz01bend,foessel11},
elliptical~\citep{ramanujan95,dbb2011ellip} or RBC-shaped
\citep{poz03rbc_shear} capsules in an unbounded shear
flow. However, in a realistic situation, biological cells and
artificial capsules are convected in bounded channels or ducts. Motivated by early experiments showing the
migration of RBCs towards the pipe centre \citep{gold1971_center_ward},
\citet{zarda77} and \cite{ozkaya87_lub} simulated the axisymmetric cellular flow
in a cylindrical tube using the finite element method
(FEM). 
Simulations based on boundary integral method (BIM), combined
with FEM for the membrane dynamics, were performed to study capsules tightly squeezed in tubes
and square ducts~\citep[e.g.][]{dbb12_pore}.
Simulations have also addressed complex phenomena like the migration and
slipper-shaped deformation of cells~\citep{poz05_nosym_tube}, suspensions of RBCs in a
capillary tube~\citep{lei2013blood}, and the shape transition between
nonaxisymmetric and axisymmetric
RBCs~\citep{mi09_ves_poi,Kaoui09_why}.
Inertial effects on the cell migration have also been
investigated numerically \citep{bagchi08_migration,shi_2d_vesicle_pous}.

These
previous computational studies focus on the capsule motion in straight
geometries. Nevertheless, capsules are seldom transported in such
simple configurations, but rather in highly complicated capillary networks as in
the \textit{in-vivo} micro-recirculation for RBCs or through
micro-fluidic devices, where corrugations, bifurcations and corners
are common. Less is known about the dynamics of capsules in
these complex geometries, although these are attracting growing
interest thanks to potential biomedical applications. Experiments~\citep{braunm11_curved_channel_11} and
simulations~\citep{noguchi_sawtooth_epl10} have shown rich behaviours
of RBCs and vesicles going through sawtooth-shaped channels; a
transition from shape oscillations to orientational oscillations was
identified for such deformable micro-objects, depending on the flow
rate and confinement. Two-dimensional FEM computations have been carried out
by~\citet{barber2008simulated} to examine the cell partitioning in
small vessel bifurcations, showing that the cells preferentially enter
the branch with higher flow rate; such an effect is intensified by the
cell migration towards the centre and hindered by 
obstructions near the bifurcations. 
~\citet{woolf11_branch} report two-dimensional simulations of a capsule in a pressure-driven channel with a side
branch. These authors found that the capsule deformation strongly depends on the branch angle and the
cells selected different paths at the branch junction according to
their deformability.
Recently, \citet{park2013transient} used the spectral boundary element method to investigate the deformation of
capsules and droplets passing through a sharp constriction in a square duct. These authors  examine
the effect of the viscosity ratio on the non-tank-treading capsule dynamics and investigate  the flow circulation
inside the capsule.

The flow passing around a corner is one of the most basic flow configurations; despite
 its universality in biological systems and micro-fluidic devices,
its influence on deformable micro-objects is not fully understood.
Steps in this direction have been taken only recently: the experiments by \cite{rusconi_corner_2010} have revealed the
rapid formation of bacterial streamers near the corners of a curved
microchannel at low Reynolds number due to the local vortical
flow structure. This secondary flow appears as long as the curvature
of the boundary varies, even in the inertialess Stokes
flow~\citep{lauga2004three}.  Simulations of an elastic
filament in a two-dimensional corner flow~\citep{2d_fila_corner_pof11} show that the filament crosses over the curved
streamlines in the corner, instead of aligning with the flow as 
in a rectilinear flow.
One of the motivations of the work is to assess whether the corner flow can be used to infer the material 
properties of soft 
particles as done by \citet{lefebvre2008flow,chu2011comparison,hu2013characterizing}. In these investigations,  
the equilibrium shape of capsules moving at a constant speed in 
confined channels or tubes is compared with that obtained from simulations or theory. As the corner 
flow is characterized by spatial non-uniformity, the capsule dynamics will undergo a transient
evolution that may therefore provide additional information on the membrane properties such as viscoelasticity.  
Knowledge of the capsule behaviour in spatially developing flows may therefore help to explore the material properties 
of soft capsules.

In this work, we numerically study the motion and deformation of an
individual capsule transported in a duct with a straight
and/or a curved corner. A three-dimensional code is developed to compute
the motion of deformable capsules in arbitrary configurations. This is based on a
boundary integral formulation with Ewald acceleration as suggested by \cite{graham07_prl}; the method  shares 
the elegance of both boundary integral and mesh-based methods. Boundary integrals
are computed to accurately account for the singular and fast-varying
interactions while the smooth part of the solution is handled by a
highly-parallel general Stokes solver based on the spectral element
method. The integration on the membrane is based on a global
spectral surface interpolation using spherical
harmonics~\citep{zhao_cell_jcp2010}. Our hybrid scheme 
couples therefore the high accuracy of boundary integrals for the short-ranged
interactions to the geometrical flexibility 
 of mesh-based methods~\citep{freund_13_review}. Spherical harmonics are utilized to resolve the membrane dynamics with spectral accuracy.
The same implementation has been used to simulate
cell sorting by deformability in a micro-fluidic device of complex geometry, i.e.\ a semi-cylindrical pillar embedded 
in a divergent channel \citep{zhu_sorting_14SM}.

The paper is organized as follows. The geometrical setup and the numerical
method are described in section~\ref{setup_method}. The results are presented in
section~\ref{results} whereas their discussion and a summary of the main
conclusions is provided in section~\ref{conclusion}.

\section{Problem setup and numerical method}\label{setup_method}
\subsection{Flow geometry and numerical procedure}

\begin{figure}
   \centering
   \includegraphics[width=0.95 \textwidth]{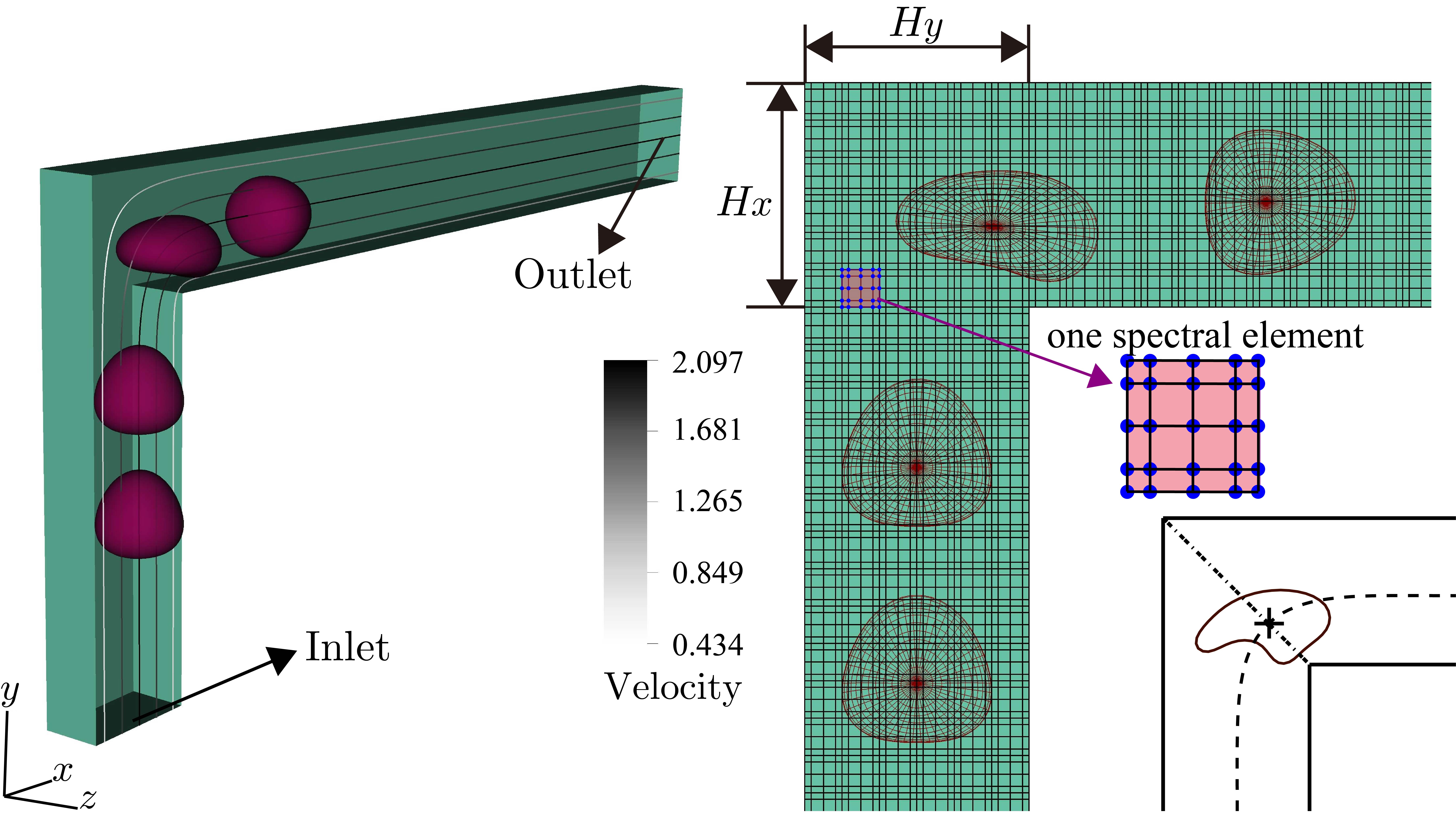}
   \caption{(Colour online) Flow configuration. Left: A
     deformable capsule travelling in a square duct with a $90$ degree straight corner,
     with half of the domain displayed. The analytical velocity profile is imposed at the
     inlet with a maximum centreline velocity of $2.097$. The flow
     field without capsules is depicted by the streaklines and the
     grey-scale colour indicating the velocity magnitude. Right: the discretized
     fluid domain and the capsule at four positions on the
     $x-y$ mid-plane. The box with blue dots represents one spectral element with
     $5\times 5 \times 5$ GLL points. Inset on the right bottom: the dashed line denotes the
     trajectory of the capsule, the dot-dashed line the
     axis of geometrical symmetry and the cross the capsule centre when located
     on the axis at instant $t=0$.}
   \label{fig:mesh_fd_cl}
\end{figure}

Figure~\ref{fig:mesh_fd_cl} displays the
flow configuration and the coordinate system used in the present investigation, where half of the flow domain
is removed to better visualize the deforming capsule. 
We investigate the motion of an elastic capsule transported through a square duct 
of width $H = H_{x} = H_{y}$; we keep $H_{x} = H_{y}$ in this work. 
In the figure, the streaklines and colour contours, coded by the velocity magnitude, are shown on the $x-y$ ($z=0$, 
omitted hereinafter)
mid-plane. The duct is characterized by a $90$ degree straight corner.

We consider an initially spherical capsule of radius $a$, enclosed by 
an infinitely thin hyperelastic membrane with surface shear modulus
$G_{s}$. The fluid inside and outside the capsule has the same density
$\rho_{F}$ and viscosity $\mu$, buoyancy forces and sedimentation effects are
neglected.

As capsules are usually small, the Reynolds number $Re$ defined with
the capsule radius $a$ and the characteristic
flow velocity $V_{\mathrm{C}}$, $Re=\rho_{F} V_{\mathrm{C}} a/\mu \ll 1$. Viscous forces are therefore 
dominant over inertial forces, and the flow inside and outside
the capsule is governed by the linear Stokes equations and determined
instantaneously by the boundary conditions. A proper tool to solve the problem
is therefore the  boundary integral method (BIM) and we adopt here an accelerated
variant of it.

The fluid-structure interaction problem is solved as follows: the flow
convects and distorts the capsule while the restoring elastic forces
alter the fluid motion~\citep{walter2010coupling}. We start with
 an undeformed capsule near the inflow and compute, 
at each time step, the elastic force on the membrane from the deformed (and out-of-equilibrium) shape of the 
capsule. 
Neglecting inertia and Brownian
fluctuations, the force density exerted by the capsule onto the fluid is equal to the membrane
load. Given this forcing, the velocities of the membrane nodes are
computed explicitly with the BIM (see the following section).

\subsection{Numerical method}\label{sec:numeric}

\subsubsection{Accelerated boundary integral method}\label{sec:ggem_bim}
We develop a boundary integral implementation accelerated by the
General Geometry Ewald like method (GGEM), proposed
by~\citet{graham07_prl} and later on used in a variety of
micro-multiphase
simulations~\citep{pranay2010pair,kumar2011seg}. An
introduction is given here, the readers are referred to
the above-mentioned articles for more details. 
The surface of the capsule $S$ is discretized by $M$ points, the Lagrangian mesh points. The elastic force per
 unit area on the membrane out of equilibrium is denoted $\bfor^{\mathrm{e}}$. 
The force per unit area from the fluid to the membrane is $\bfor^{\mathrm{f}}$,
with $\bfor^{\mathrm{f}} + \bfor^{\mathrm{e}} = 0$ due to the stress continuity.
In return, the force per unit volume exerted by the deforming surface onto the fluid at position $\bx$ is
$\brho \bxpp = \int_{S}-\bfor^{\mathrm{f}} \delta \lp \bx -\bxm \rp  dS\lp \bxm \rp = \int_{S}\bfor^{\mathrm{e}}
\delta \lp \bx -\bxm \rp  dS\lp \bxm \rp$, with $\delta$ the Dirac delta
function. We thus need to solve the following equations for the fluid in the 
inertialess Stokes regime
\begin{align} \label{eq:stokes}
-\boldsymbol{\nabla}p\bxpp + \mu\nabla^{2}\bu\bxpp + \int_{S}\bfe \lp \bxm \rp \delta \lp \bx -\bxm
\rp  dS\lp \bxm \rp = 0,\nonumber \\ 
\boldsymbol{\nabla} \cdot \bu \bxpp = 0,
\end{align}
where $p$ and $\mathbf{u}$ denote the pressure and fluid velocity and $\mu$ is the
fluid dynamic viscosity. Owing to the linearity of the Stokes problem, the flow field
can be expressed as a boundary integral on the surface of the capsule only,
\begin{equation}
 \bubx = \bu^{\infty}\lp \bx\rp + \int_{S} \mathbf{G}\lp \bx,\bxm \rp\cdot \bfe \lp \bxm \rp dS\lp \bxm \rp,
\end{equation}
where $\bu^{\infty}\left(\bx\right)$ is the velocity field of the undisturbed flow and
$\mathbf{G}\left(\mathbf{x}^{\prime},\mathbf{y}^{\prime}\right)$ is the free-space Green's
function of the Stokes problem, also known as the Stokeslet or \textit{Oseen-Burgers} tensor,
\begin{equation} \label{eq:greenfun}
\mathbf{G}\left(\mathbf{x}^{\prime},\mathbf{y}^{\prime}\right) = \frac{1}{8\pi\mu
r}\lp\boldsymbol{\delta}+\frac{\left( \xmy \right) \left( \xmy \right)}{r^2}\rp,
\end{equation}
with $r=|\xmy|$.

The GGEM method decomposes the force per unit volume $\brho \bxpp$ into a local part
$\brhol \bxpp$ and a global part $\brhog \bxpp$, with $\brho \bxpp = \brhol \bxpp + \brhog \bxpp$
and
\begin{align}\label{eq:global_forcing}
\brhol \bxpp &= \int_{S}\bfe \bxmpp [ \delta - g\lp \bx - \bxm \rp ] dS \bxmpp,\nonumber\\
\brhog \bxpp &= \int_{S}\bfe \bxmpp  g\lp \bx - \bxm \rp  dS \bxmpp,
\end{align}
where $g\lp\bx^{\prime}\rp$ is a quasi-Gaussian function used to smoothen the Dirac delta function,
\begin{align}
g\left(\mathbf{x}^{\prime}\right)=\left(\alc^3/\pi^{3/2}\right)e^{\left(-\alc^
{2}|\mathbf{x}^{\prime}|^{2}\right)}\left[5/2-\alc^{2}|\mathbf{x}^{\prime}|^{2}\right],
\end{align}
where $\alc^{-1}$ indicates the length scale over which the smoothing is active.

The Stokes problem in Eq.~\ref{eq:stokes} is therefore decomposed into two
problems: one for the flow induced by the local force $\brhol \bxpp$ hence called the local problem and one for its 
global counterpart  $\brhog \bxpp$.
The velocity field $\bubx$ is the sum of the local velocity $\bul\bxpp$ and global velocity
$\bug\bxpp$, $\bubx=\bul\bxpp+\bug\bxpp$. The local
problem accounts for the singular and short-ranged interaction while the global problem for the smooth and long-ranged
interactions. These are solved by different numerical methods; the local solution is
calculated by the boundary integral method due to its superior accuracy in resolving fast-decaying interactions, while
the global problem is handled by a mesh-based solver that provides geometrical flexibility. 

The modified Stokeslet pertaining the local problem can be shown to be  
\begin{equation}\label{eq:mod_green}
\bG^{\mathrm{l}}\lp \bxprm
\rp=\frac{1}{8\pi\mu}\lp\bdel+\frac{\bxprm \bxprm}{|\bxprm|^2}\rp\frac{\mathrm{erfc}
\lp\alc |\bxprm|\rp}{|\bxprm|}-\frac{1}{8\pi\mu}\lp\bdel-\frac{\bxprm
\bxprm}{|\bxprm|^2}\rp\frac{2\alc}{\pi^{1/2}}e^{\lp-\alc^{2}  |\bxprm|^{2}\rp},
\end{equation}
so that the velocity field $\bu^{\mathrm{l}} \bxpp$ of the local solution can be obtained as
\begin{equation}~\label{eq:velloc_int}
 \bu^{\mathrm{l}} \bxpp = \int_{S} \bG^{\mathrm{l}}\lp \bx,\bxm \rp\cdot \bfe \lp \bxm \rp dS\lp \bxm
\rp.
\end{equation}
\sloppy
Eq.~(\ref{eq:velloc_int}) can be integrated by classical boundary integral implementations. Regularised Stokeslets can be used to
facilitate the calculations, as done
among others in~\citet{graham07_prl,pranay2010pair}. Nonetheless, the BIM with
regularisation suffers a degradation of the numerical accuracy and
robustness for cases involving strong confinement or closely packed
objects.
Singular and nearly-singular integration is necessary 
to achieve the required accuracy in these cases~\citep{curse_nearsing,zhu2013low}, 
and this is the approach pursued here.

The modified Stokeslet $G^{\mathrm{l}} \lp \bx^{\prime} \rp $ is valid for an unbounded domain, thus the
local velocity $\bu^{\mathrm{l}} \bxpp$ does not account for the influence of any additional boundaries. 
The global velocity will therefore be defined in such a way that the sum of the two will satisfy the 
required boundary conditions, no slip at the solid wall $\Omega$ in the cases investigated here, $\bu^{\mathrm{l}} \lp 
\bx_{\Omega} \rp + \bu^{\mathrm{g}} \lp \bx_{\Omega} \rp = 0$.
The global problem amounts to solving the Stokes problem in the domain of interest with the known volume forcing $\brhog\bxpp$
and boundary conditions defined by the solution of the local problem. This allows the use of a variety of
efficient and accurate numerical methods
for the solution of the Stokes equations in any complex geometry. Here, we compute the global solution with 
Stokes module of the open-source Navier-Stokes
solver NEK5000~\citep{nek5000-web-page}, using the spectral element method. NEK5000 has been extensively used for
stability analysis~\citep{schrader2010receptivity} and
turbulent flows~\citep{fischer2008petascale} in complex domains. As akin to FEM, the physical domain is decomposed
into elements with each element subdivided into arrays of
Gauss-Lobatto-Legendre (GLL) nodes for the velocity and Gauss-Legendre
(GL) nodes for the pressure field. The Galerkin approximation is 
employed for the spatial discretization with different velocity and
pressure spaces, the so-called
$\mathbb{P}_{N}-\mathbb{P}_{N-2}$
approach~\citep{maday1989spectral}. Accordingly, the velocity
(respectively pressure) space consists of $N\mathrm{th}$
(respectively ~$\left(N-2\right)\mathrm{th}$) order Lagrange polynomial
interpolants, defined on the GLL (respectively GL) quadrature points in each
element. Note that we do not solve the Navier-Stokes equations with a very small but finite
Reynolds number, but instead use the steady Stokes solver of NEK5000 at each time step.
NEK5000 is chosen here for its spectral accuracy, high parallel performance and most importantly
its geometric flexibility, fully exploiting the general-geometry merit
of GGEM.

The global problem is solved only on the Eulerian mesh points, which do not necessarily coincide with the Lagrangian
mesh points on the membrane (see Fig.~\ref{fig:mesh_fd_cl}). Thus, at each time step, an interpolation from the
global solution is performed to obtain the global velocity $\bug \lp \bx_{\mathrm{i}}\rp, \mathrm{i}=1,2,3,...M$ of the
Lagrangian points. The interpolation error is minimized thanks to the spectral accuracy of NEK5000. 
The velocities of the Lagrangian points are obtained by summing up the local and global velocities. We use
a third-order Adam-Bashforth time-integration scheme to update the position of those points.

In our work, we choose $\rcut=4\alc^{-1}$ and
$\alc=a^{-1}$ as in the work of~\citet{pranay2010pair}. The alternative value $\rcut=5\alc^{-1}$ has also been tested
for some of the cases and no significant differences have been observed.

\subsubsection{Spectral method for the membrane dynamics}\label{sec:membrane_spectral}

The membrane loading was calculated as linear piece-wise functions on
triangular meshes by~\citet{poz95shear,ramanujan95,li2008front} among others. 
FEM has been also implemented by \citet{walter2010coupling} for its generality and versatility.
Bi-cubic B-splines
interpolation functions are adopted by~\citet{lac2007_bspl} to
obtain accurate results at a reasonably high computational
cost. Alternatively, an accurate spectral boundary element algorithm is used
by~\citet{dimitra2009_extention,dimitra11_squre, kuriakose2013deformation}, thus coupling
the numerical accuracy of the spectral method and the geometric
flexibility of the boundary element method. Another attractive alternative
is the global spectral method. Fourier spectral interpolation and spherical
harmonics are used
for two-dimensional \citep{freund2007leukocyte} and three-dimensional
simulations \citep{kessler08_global_spectral,zhao_cell_jcp2010}. Here, we follow the approach of \citet{zhao_cell_jcp2010}, briefly outlined 
below.

We map the capsule surface onto the surface of the unit reference sphere $\mathbb{S}^{2}$, using its
spherical angles $\left(\theta,\phi\right)$ for the
parametrisation. The parameter space $\left\{\left(\theta,\phi\right)
|0 \leqslant \theta \leqslant \pi, 0 \leqslant \phi \leqslant
2\pi\right\}$ is discretized by a quadrilateral grid consisting of
Gauss-Legendre quadrature points in $\theta$ and uniform intervals in
$\phi$. 
All other surface quantities are defined on the same mesh. The
surface coordinates $\mathbf{x}\left(\theta,\phi\right) $ are
expressed by a truncated series of spherical harmonic functions,
\begin{equation} 
\bx\lp\theta,\phi\rp =
\sum_{n=0}^{\NS}\sum_{m=0}^{n}\bar{P}_{n}^{m}\lp \cos \theta\rp \lp
\mathbf{a}_{nm}\cos m\phi + \mathbf{b}_{nm}\sin m\phi\rp ,
\end{equation}
 yielding $N_{\mathrm{SH}}^{2}$ spherical harmonic
 modes. The corresponding normalised  Legendre polynomials are
\begin{equation}
\bar{P}^{m}_{n}\left(x\right)=\frac{1}{2^{n}n!}\sqrt{\frac{
\left(2n+1\right)\left(n-m\right)!}{2\left(n+m\right)!}}\left(1-x^{2}\right)^{
\frac{m}{2}}\frac{d^{n+m}}{dx^{n+m}}\left(x^2-1\right)^{n}.
\end{equation}
Both forward and backward transformations are calculated with the
SPHEREPACK library~\citep{adams1997spherepack,swarztrauber2000generalized}.
Aliasing errors arise due to the nonlinearities induced by the membrane model
and the complicated geometry (products, roots and inverse operations needed to calculate the geometric quantities 
introduced below). 
We implement an approximate dealiasing by performing the nonlinear
operations on $M_{\mathrm{SH}}>N_{\mathrm{SH}}$ points and filtering the result back to $N_{\mathrm{SH}}$
points.
A detailed discussion on this issue is provided in \citet{freund+zhao10}.

A point on the surface is expressed by the curvilinear coordinates,
$\lp \xi^{1},\xi^{2} \rp = \left(\theta,\phi\right)$, defined on the covariant base,
$\left(\mathbf{a}_{1},\mathbf{a}_{2},\mathbf{a}_{3}\right)$, following
the local deformation. The base vectors are

\begin{align}
  \mathbf{a}_{1}  = \frac{\partial \mathbf{x}}{\partial \theta},
  \mathbf{a}_{2}  = \frac{\partial \mathbf{x}}{\partial \phi},
  \mathbf{a}_{3}  =\mathbf{n} =
\frac{\mathbf{a}_{1}\times\mathbf{a}_{2}}{|\mathbf{a}_{1}\times\mathbf{a}_{2}|},
\end{align}
and the covariant and contravariant metric tensors
\begin{align}
a_{\alpha \beta} = \mathbf{a}_{\alpha} \cdot \mathbf{a}_{\beta},
a^{\alpha \beta} = \mathbf{a}^{\alpha} \cdot \mathbf{a}^{\beta},
\end{align}
where $\alpha,\beta=1,2$.
The base vectors and metric tensors are also defined for the undeformed state and denoted here by 
capital letters ($\mathbf{A}^{\alpha}$, $A^{\alpha \beta}$).

The second fundamental form coefficient of the surface is $b_{\alpha\beta}=\mathbf{n} \cdot
\frac{ \partial \mathbf{a}_{\alpha}}{\partial \xi^{\beta}}$ and the two invariants of the transformation $I_{1}$ and 
$I_{2}$
are
defined as
\begin{equation}
I_{1}=A^{\alpha\beta}a_{\alpha\beta}-2,
I_{2}=|A^{\alpha\beta}||a_{\alpha\beta}|-1.
\end{equation}
$I_{1}$ and $I_{2}$ can also be determined from the principal
dilations $\lambda_{1}$ and $\lambda_{2}$,
\begin{equation}
I_{1}=\lambda_{1}^{2}+\lambda_{2}^{2}-2,I_{2}=\lambda_{1}^{2}\lambda_{2}^{2}
-1=J_{2}^{2}-1.
\end{equation}
The Jacobian, $J_{s}=\lambda_{1}\lambda_{2}$, shows the ratio of the
deformed to the undeformed surface area.
We compute the in-plane Cauchy stress tensor
$\mathbf{T}$, from the strain energy function per unit area of the
undeformed membrane, $W_{S}\left(I_{1},I_{2}\right)$,
\begin{equation}~\label{eq:cauchy_tensor}
\mathbf{T}=\frac{1}{J_{s}}\mathbf{F}\cdot\frac{\partial W_{S}}{\partial
\mathbf{e}}\cdot \mathbf{F}^{T},
\end{equation}
where $\mathbf{F}$ is $\mathbf{a}_{\alpha} \otimes
\mathbf{A}^{\alpha}$. Eq.~(\ref{eq:cauchy_tensor}) can
be further expressed by components as
\begin{equation}
  T^{\alpha\beta}=\frac{2}{J_{s}}\frac{\partial W_{S}}{\partial
I_{1}}A^{\alpha\beta}+2J_{s}\frac{\partial W_{S}}{\partial
I_{2}}a^{\alpha\beta}.
\end{equation}
We employ a widely-used model of the strain energy function $W_{S}$ in our
study, the neo-Hookean law (NH)~\citep{large_deform} formulated as 
\begin{eqnarray}
W_{S}^{\mathrm{NH}} & = & \frac{G_{s}}{2}\left(I_{1}-1+\frac{1}{I_{2}+1}\right),
\end{eqnarray}
where $G_{s}$ is the surface shear modulus. The local equilibrium
connects $\mathbf{T}$ with the external membrane load
$\mathbf{q}$, as
\begin{equation}\label{eq:local_equ}
  \nabla_{s} \cdot \mathbf{T} + \mathbf{q} = 0,
\end{equation}
where ($\nabla_{s}\cdot$) is the surface divergence operator in the deformed
state. In curvilinear coordinates, the load vector is written as $\mathbf{q} =
q^{\beta}\mathbf{a}_{\beta}+q^{n}\mathbf{n}, \; \beta=1,2$.
The local balance in Eq.~(\ref{eq:local_equ}) is further
decomposed into tangential and normal components,
\begin{eqnarray}\label{eq:loca_equ_com}
\frac{\partial T^{\alpha \beta}}{\partial \xi^{\alpha}} +
\Gamma^{\alpha}_{\alpha\lambda}T^{\lambda \beta} +
\Gamma^{\beta}_{\alpha\lambda}T^{\alpha\lambda} +q^{\beta}& = & 0, \quad
\beta = 1, 2, \nonumber \\ T^{\alpha\beta}b_{\alpha\beta} + q^{n} &=&
0,
\end{eqnarray}
where $\Gamma_{\alpha\lambda}^{\beta}$ are the Christoffel symbols.

We incorporate bending stiffness into our model using the linear
isotropic model for the bending moment $\mathbf{M}$:
$M^{\alpha}_{\beta}=-G_{B}\left(b^{\alpha}_{\beta}-B^{\alpha}_{\beta}\right)$,
where $G_{B}$ is the bending modulus, and $b^{\alpha}_{\beta}$ is the mixed version of the second fundamental form
coefficients ($B^{\alpha}_{\beta}$ corresponds to that of the reference configuration). 
Considering the local torque balance with bending moments exerted on the membrane, we obtain the transverse shear 
vector $\bQ$ and in-plane stress tensor $\bT$, 
\begin{align}
 M^{\alpha\beta}_{|\alpha}-Q^{\beta} =0 \label{eq:tran_bend},\\
 \varepsilon_{\alpha\beta}\lp T^{\alpha\beta}-b^{\alpha}_{\gamma}M^{\gamma\beta} \rp = 0,\label{eq:inplan_bend}
\end{align}
where '$_{|\alpha}$' denotes the covariant derivative and $\boldsymbol{\varepsilon}$  the 
two-dimensional Levi-Civita tensor. Eq.~(\ref{eq:inplan_bend}) determines
the antisymmetric part of the in-plane stress tensor, which is always zero as proved in \cite{zhao_cell_jcp2010}. 
Including the 
transverse shear stress
$\bQ$, the local  
equilibrium of the stress, including bending, gives
\begin{eqnarray}\label{eq:loca_equ_bend}
\frac{\partial T^{\alpha \beta}}{\partial \xi^{\alpha}} +
\Gamma^{\alpha}_{\alpha\lambda}T^{\lambda \beta} +
\Gamma^{\beta}_{\alpha\lambda}T^{\alpha\lambda}
-b^{\beta}_{\alpha}Q^{\alpha} + q^{\beta}& = & 0, \beta = 1, 2,
\nonumber \\ T^{\alpha\beta}b_{\alpha\beta} + Q_{|\alpha}^{\alpha} +
q^{n} &=& 0.
\end{eqnarray}

\subsubsection{Singular and nearly-singular integration}
In this section, we report the scheme for singular and nearly-singular integration based on the spectral
surface discretization. We mostly follow the approach in \citet{zhao_cell_jcp2010}, which
is shortly described here for the sake of completeness.
We rewrite the boundary integral equation
Eq.~(\ref{eq:velloc_int}) in its general form as
\begin{align}\label{eq:int_sph}
 \mathcal{I}\lp \bxzero \rp = \int_{S} K\lp \bx,\bxzero \rp g\bxpp d S\bxpp = \int_{\mathbb{S}^{2}} K \lp \bx \lp 
\theta,\phi
\rp,\bxzero \rp g \lp \bx \lp \theta,\phi \rp \rp J \lp \theta,\phi \rp d\theta d\phi,
\end{align}
where $K$ is one component of the Green's function kernel, the modified Stokeslet in Eq.~(\ref{eq:mod_green}) in our
case, $g$ is a smooth function defined in $S$ and $J=|\frac{\partial \bx}{ \partial \theta}
\times \frac{\partial \bx}{ \partial \phi}|$ the Jacobian. If the point $\bxzero$ is sufficiently far
from
the membrane 
surface $S$, $K$ is smooth  and the integral $\mathcal{I}\lp \bxzero \rp$ can be computed as 
\begin{equation}~\label{eq:integ}
  \mathcal{I} \lp \bxzero \rp = \sum_{k=1}^{M = \NS \times  2\NS } K\lp \bx_{k},\bxzero \rp g\lp \bx_{k} \rp J\lp
  \theta_{k}, \phi_{k}\rp 
\omega_{k}, 
\end{equation}
where $\omega_{k}$ is the weight of the $k$th discretized point. If $\bxzero$ lies on the boundary 
$S$, the kernel function $K\lp \bx,\bxzero \rp $ becomes singular: in this case, a naive
integration using Eq.(\ref{eq:integ}) would give low accuracy; this so-called singular integration needs a
special treatment. As $\bxzero$ is very close to $S$, $K\lp \bx,\bxzero \rp $ becomes nearly-singular, also requiring
additional care. We adopt here the approach denoted as floating partition
of unity~\citep{bruno2001fast}. 
In the singular case, $\bxzero$ on the surface, we define $s\lp
\bx,\bxzero
\rp $ as the contour length along the great circle connecting $\bx$ and $\bxzero$ on the reference
sphere $\mathbb{S}^{2}$. This is
used to define a mask function $\eta \lp s\lp \bx,\bxzero \rp \rp$,
\begin{equation}
\eta \lp s \rp = \left\{ \,
\begin{IEEEeqnarraybox}[][c]{l?s}
\IEEEstrut
\exp{\lp \frac{2 \exp{\lp-1/t\rp}}{t-1}\rp} & if $t = s/s_{\mathrm{cut}} < 1$, \\
0 & if $s \geqslant s_{\mathrm{cut}}$,
\IEEEstrut
\end{IEEEeqnarraybox}
\right.
\label{eq:maskfun}
\end{equation}
where $s_{\mathrm{cut}}$ is a cut-off radius. With the mask function $\eta \lp s \rp$, the boundary integral $\mathcal{I}\lp
\bxzero \rp$ is decomposed into two parts, a singular part $\mathcal{I}_{\mathrm{singular}}\lp
\bxzero \rp$ and a smooth part
$\mathcal{I}_{\mathrm{smooth}}\lp
\bxzero \rp$,
\begin{align}\label{eq:int_unit}
 \mathcal{I}\lp \bxzero \rp & =  \mathcal{I}_{\mathrm{singular}}\lp \bxzero \rp + \mathcal{I}_{\mathrm{smooth}}\lp 
\bxzero \rp,\\ \nonumber
 \mathcal{I}_{\mathrm{singular}}\lp \bxzero \rp & =  \int_{S} \eta\lp s\lp \bx,\bxzero  \rp \rp K\lp  
\bx,\bxzero \rp
g\bxpp d S\bxpp ,\\ \nonumber
\mathcal{I}_{\mathrm{smooth}}\lp \bxzero \rp & =    \int_{S} \left[ 1 -\eta\lp s\lp \bx,\bxzero  \rp \rp \right ] K\lp 
\bx,\bxzero \rp
g\bxpp d S\bxpp.
\end{align}
The integrand of the smooth part becomes zero as $\bx$ and $\bxzero$ coincide so that the integral can be computed accurately
using Eq.~(\ref{eq:integ}). The singular part has non-zero values only on
the spherical patch of radius $s_{\mathrm{cut}}$, and it
can be integrated using local polar coordinates defined on that patch,
\begin{equation}\label{eq:int_singu}
 \mathcal{I}_{\mathrm{singular}}\lp \bxzero \rp=\int_{0}^{2\pi}\int_{0}^{s_{\mathrm{cut}}}\eta\lp s
\rp K\lp \bx \lp s, \psi \rp,\bxzero \rp g\lp s,  \psi \rp J^{\prime}\lp s,\psi  \rp ds d\psi,
\end{equation}
where $J^{\prime}\lp s,\psi  \rp=|\frac{\partial \bx}{ \partial s}
\times \frac{\partial \bx}{ \partial \psi}|$ is the Jacobian of the transformation. We apply Gauss
quadrature along the radial direction $s \in [0,s_{\mathrm{cut}}]$ and sum over the 
circumferential direction $\psi \in [0, 2\pi]$. The radius of the patch is chosen to be $s_{\mathrm{cut}} =
\pi/\sqrt{\NS}$ \cite[see the detailed discussion in][]{zhao_cell_jcp2010}. Because the
quadrature points do not necessarily coincide with the discretization points, interpolation is needed to obtain
quantities such as $g\lp s, \psi \rp$. Bi-cubic spline interpolation is performed here: firstly, we
compute $g$ on a uniform mesh in $\theta$ and $\phi$ based on the spherical harmonic coefficients; the mesh
is then extended from $\theta \in [0,\pi]$ to $\theta \in [0,2\pi]$ exploiting the symmetry $g\lp 2\pi-\theta,\pi+\phi
\rp = g\lp \theta, \phi\rp$; $g$ is periodic in both directions on the extended domain and its derivatives can
be accurately computed by Fourier transform; we finally construct the bi-cubic spline approximation using the function
derivatives.

For the nearly-singular integration, we first find the projection of $\bxzero$ onto the membrane
surface, $\bxzero^{\mathrm{proj}}$, and then compute the boundary integral on the spherical patch centred
at $\bxzero^{\mathrm{proj}}$. A $\sinh$ transformation is applied in the radial direction in order to move the
quadrature points closer to $\bxzero^{\mathrm{proj}}$~\citep{johnston2005sinh}, better resolving the fast-varying
Green's function near $\bxzero^{\mathrm{proj}}$.

\subsection{Nondimensionalization}
The capsule membrane is characterised by
its resistance to shearing and bending. The capillary number $Ca$, the
ratio of viscous over elastic forces, is defined based on the surface shear modulus $G_{s}$,
\begin{equation}
  Ca=\frac{\mu V_{\mathrm{C}}}{G_{s}},
\end{equation}
where we use the mean velocity as the characteristic flow velocity $V_{\mathrm{C}}$. The reduced bending modulus,
$Cb$, is the ratio of the bending and shearing moduli,
$Cb=G_{B}/a^2G_{s}$. We use the radius of the capsule $a$ as 
the reference length scale, so that the characteristic time scale is $T=a/V_{\mathrm{C}}$.

\subsection{Validation}
We firstly introduce the parameters used in the discretization. 
As mentioned in section~\ref{sec:ggem_bim}, 
$\alc=a^{-1}=1$ and $\rcut=4\alc^{-1}$ are adopted
following the recommendation in \citet{pranay2010pair}.
Cubic spectral elements of size $1$ with $5\times 5 \times 5$ GLL points are used to
discretize the fluid domain, where the mean grid spacing $h_{\mathrm{mean}}=1/4$ 
well satisfies the relation $\alc h_{\mathrm{mean}} \leq 0.5$ proposed in 
\citet{graham12_jcp}. Rigorous tests are carried out to be sure that the results are independent of
the mesh resolution and the cut-off radius $\rcut$, supporting the current choice.
For the membrane dynamics, $N_{\mathrm{SH}}=24$ modes with
a dealiasing factor $M_{\mathrm{SH}}/N_{\mathrm{SH}}=2$ are chosen 
to exploit the geometrical symmetry.

\begin{figure}
    \centering
    \subfigure[]{\label{fig:neo_valid_nobend}
    \includegraphics[scale = 0.16] {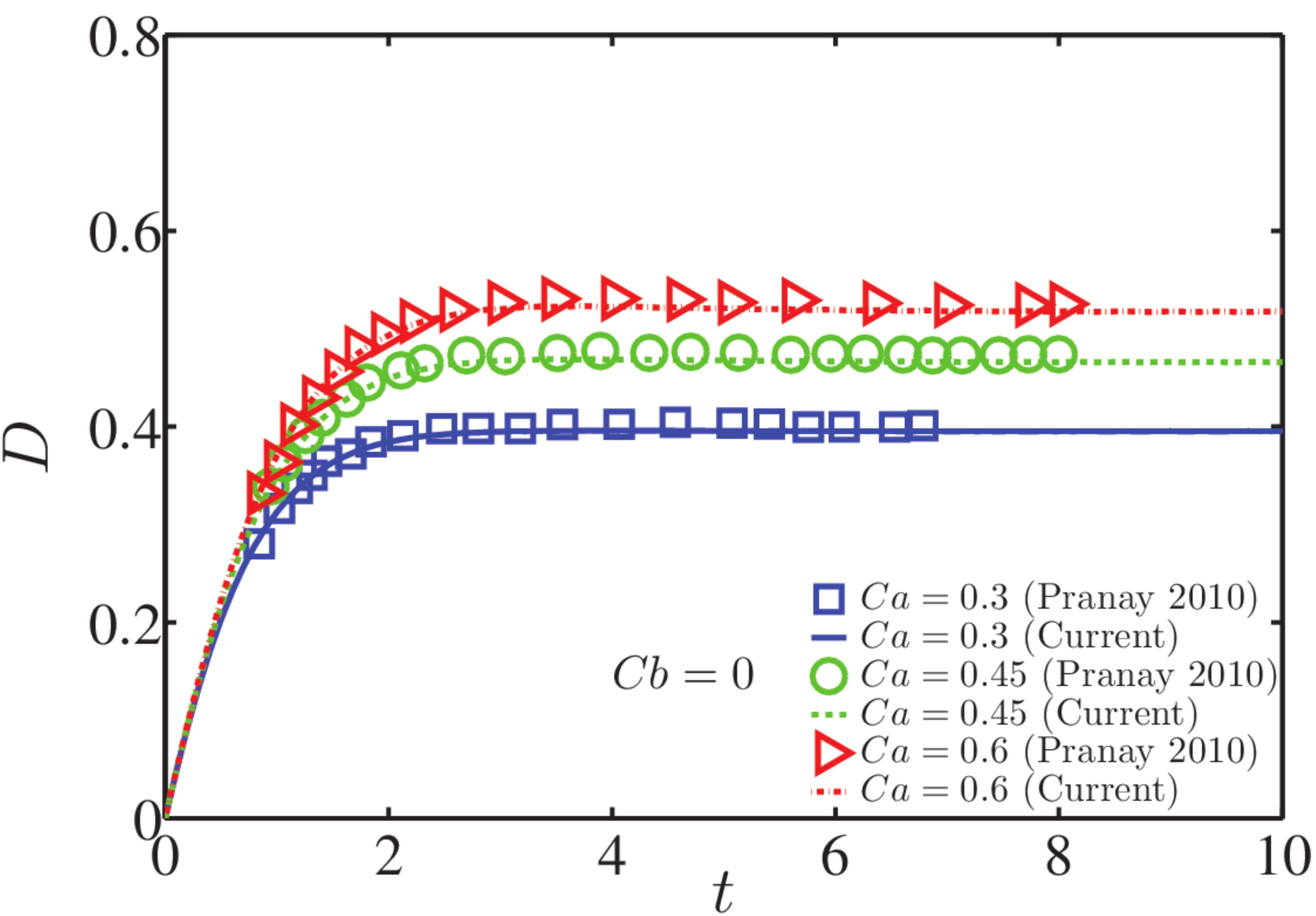}\hspace{0em}
 }
    \subfigure[]{\label{fig:neo_valid_bend}
    \includegraphics[scale = 0.16] {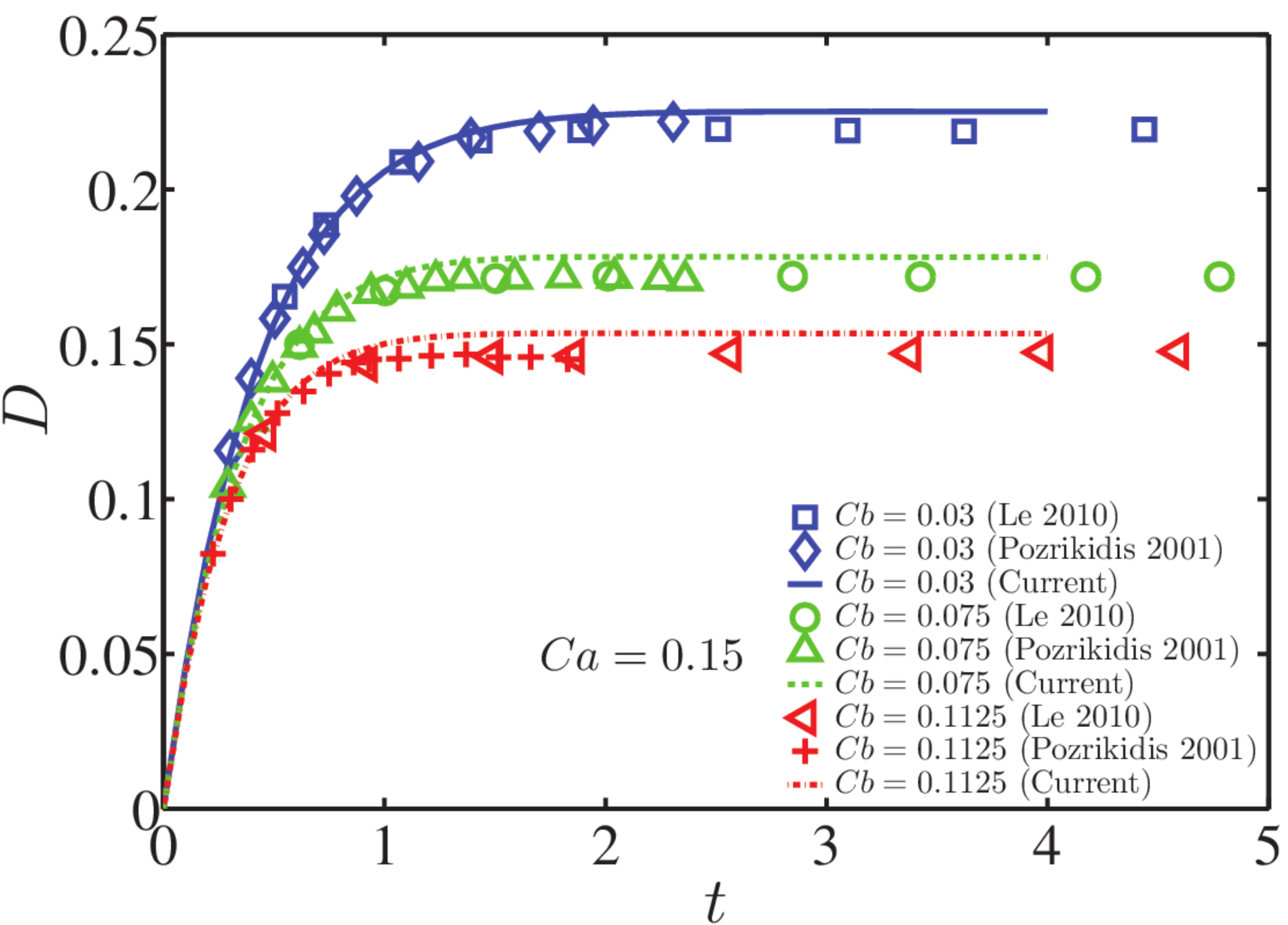}
 }
    \caption{(Colour online) \subref{fig:neo_valid_nobend} Variation of
     the deformation parameter $D$ versus time for an initially
     spherical neo-Hookean capsule in shear flow. Different values of the capillary number $Ca$ are
     chosen. The profile of the capsule in the shear plane is an ellipse
     with a long axis $L_{\mathrm{max}}$ and short axis
     $L_{\mathrm{main}}$; the Taylor parameter quantifying the capsule
     deformation is $D=\frac{L_{\mathrm{max}}-L_{\mathrm{min}}}{L_{\mathrm{max}}+L_{\mathrm{min}}}$. 
     \subref{fig:neo_valid_bend} Same
     as figure~\subref{fig:neo_valid_nobend}, with $Ca=0.15$ and different values of the reduced bending modulus $Cb$.}
   \label{fig:neo_valie}
 \end{figure}

The tank-treading motion of an initially spherical capsule in homogeneous shear flow is
selected as the first validation case of our implementation. The capsule 
evolves into a prolate and reaches a steady deformed shape where the membrane continuously
rotates in a tank-treading fashion. The time-dependent capsule deformation is measured
by the Taylor parameter
\begin{equation}
D=\frac{L_{\mathrm{max}}-L_{\mathrm{min}}}{L_{\mathrm{max}}+L_{\mathrm{min}}},
\end{equation}
where $L_{\mathrm{max}}$ and $L_{\mathrm{min}}$ are the maximum and minimum dimensions
of the capsule in the shear plane. We display $D$ as a function of time for neo-Hookean
capsules with a varying $Ca$ and no bending stiffness in figure~\ref{fig:neo_valid_nobend}. Good
agreement is observed between our simulations and those of \citet{pranay2010pair}. 

\begin{figure}
    \centering
    \includegraphics[scale = 0.3] {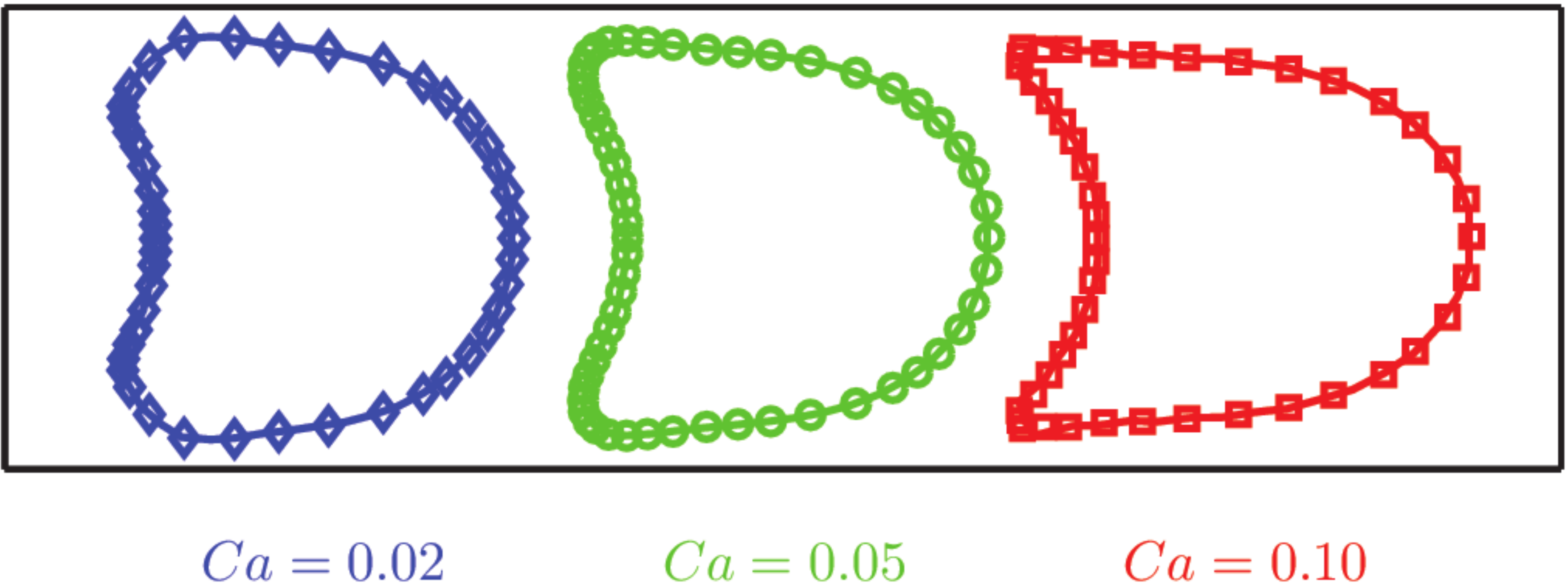}
   \caption{(Colour online) Equilibrium profiles of neo-Hookean
     capsules with different capillary number $Ca$ in a square duct of
     size $l_{\mathrm{duct}}$ and confinement
     $2a/l_{\mathrm{duct}}=0.9$, $Cb=0$. The symbols correspond to the
     results of~\citet{dbb12_pore} and the solid lines to our simulations using $N_{\mathrm{SH}}=24$ modes to represent 
the membrane surface.}
     \label{fig:neo_duct_pore}
 \end{figure}

We next compare cases including bending modulus against the 
results of ~\citet{poz01bend}
and~\citet{le10bend}, see figure~\ref{fig:neo_valid_bend}. The agreement is generally good although small 
differences appear when the capsule reaches its
equilibrium shape. This is probably due to the different discretization used to evaluate the high-order derivatives
for the calculation of bending moments. 
 As pointed out by
\citet{poz01bend}, his simulations suffer from 'significant inaccuracies' at high capsule deformations; our results
agree very well with theirs in the small deformation regime (around $t<0.5$).
To verify the nearly-singular integration,
we therefore also simulate a capsule with zero bending stiffness compressed in a confined square duct, and report 
excellent agreement with the data of \citet{dbb12_pore}, see figure~\ref{fig:neo_duct_pore}.

\section{Results}\label{results}
We consider an initially spherical capsule located at  the centre of the square duct, as
deformable objects tend to move towards the centreline due to the
F\r{a}hraeus effect. 
We impose the analytical velocity profile of a rectangular duct 
flow~\citep{spiga1994symmetric} at the inlet with  mean velocity  $V_{\mathrm{C}}$.
We anchor the centre of the capsule at 
$\left(0,-5,0\right)a$, i.e.\ $5a$ away from both the computational inlet and the corner,
and release it after it has reached its equilibrium shape. This distance
is large enough for the interaction between the capsule and the inlet/corner
to be negligible during this initial phase.

We investigate the influence of the capillary number $Ca$ on the dynamics of the capsule,
including its deformation, trajectory, velocity, surface area and principal tensions.
The reduced bending modulus is fixed to $Cb=0.04$, unless otherwise specified. 
In addition, we examine the influence of the confinement and of the geometry of the corner.

We note $Cb \approx 
0.01$ for RBCs, according to ~\citet{poz_file_RBC_pof,zhao_cell_jcp2010}. We adopt the larger value 
$Cb=0.04$ to prevent the bulking of membrane that would easily destabilize the simulations. Luckily, we found the 
influence by varying 
$Cb$ is much weaker than that by varying the capillary number $Ca$. Indeed, 
$Cb$ represents the relative strength of bending over shearing and its variation from $0.01$ to $0.04$ 
accounts for only $3\%$ of the shear modulus.

\subsection{Square duct flow with a $90^{\circ}$ straight corner} We begin by investigating the motion of a 
capsule transported in a moderately confined square duct (of width $H_{x}=3a$) with a straight corner.
Throughout the work, the cross section of the vertical and horizontal duct remains 
the same, $H_{y} \equiv H_{x}$.

\begin{figure}
    \centering
    \includegraphics[trim=0cm 0cm 0cm 4cm, clip=true, width = 110mm] {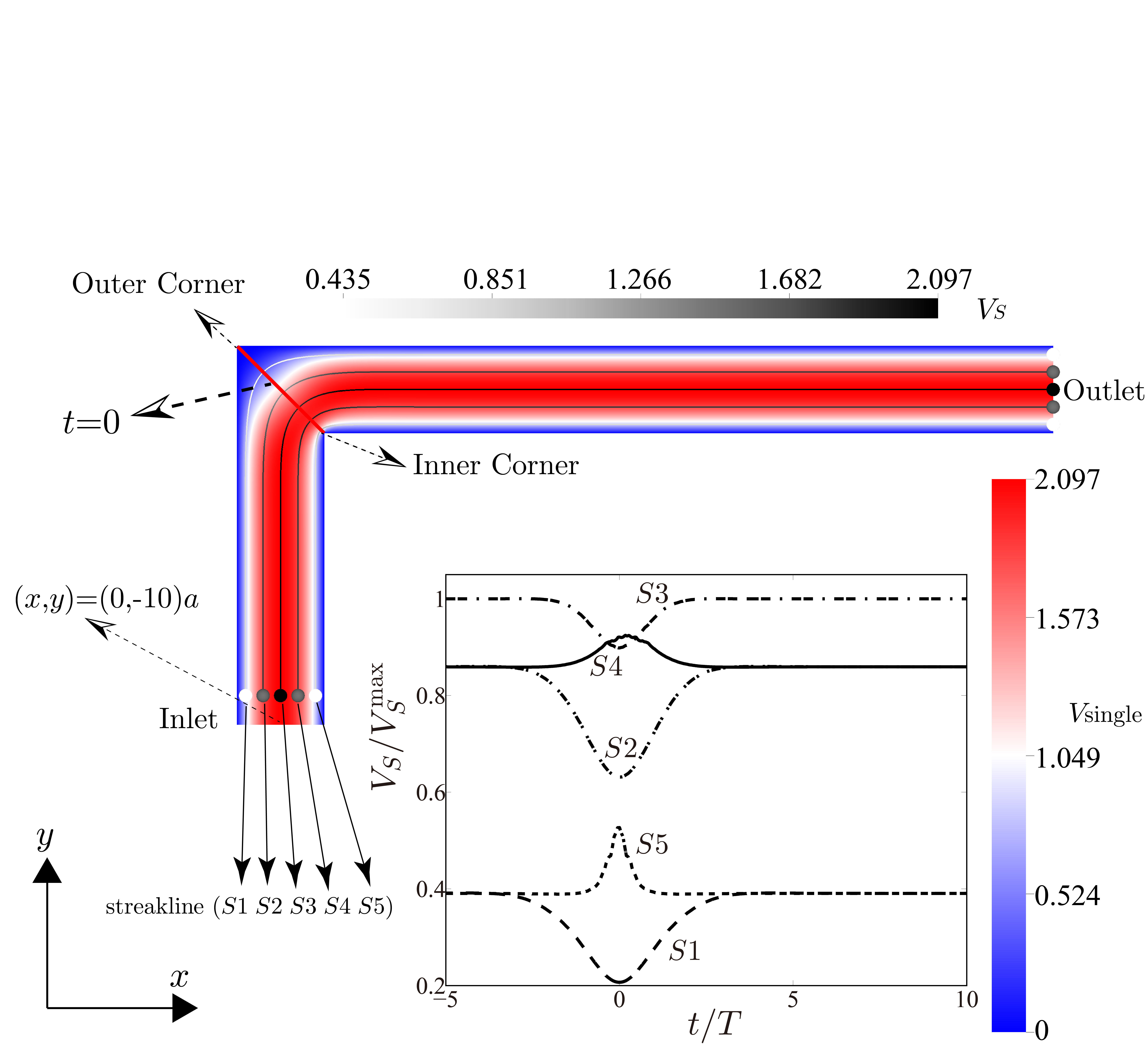}
    \caption{(Colour online) The velocity field pertaining the
      single-phase flow in a square duct
      of width $H_{x}=3a$ with a straight corner. 
      The flow field and streaklines are coloured by
      their magnitude $V_{\mathrm{single}}$ and $V_{S}$ respectively.
      The streaklines $(S1,S2,S3,S4,S5)$ start from the
      points equally spaced between $(-1.2,-9)a$ and
      $(1.2,-9)a$. 
       $V_{S}$ divided by the maximum flow velocity is depicted in the inset
      versus time, where $t=0$ corresponds to the time when a fluid particle
      crosses the corner symmetry axis.}
     \label{fig:streamline_background}
 \end{figure}

The background flow in the absence of capsules is refereed to as the
single-phase flow and is illustrated
in figure~\ref{fig:streamline_background}  in the $x-y$  plane. We show five trajectories
$\left(S1,S2,S3,S4,S5\right)$ starting from equally-spaced points
on the line $y=-9a,\;x \in [-1.2,1.2]a$; they are ordered from the outer to the inner corner so that $S3$ goes through 
the centre of the domain. The velocity
magnitude $V_{S}\left(t\right)$ is symmetric about $t=0$, when the minimum
is reached for $S1$, $S2$ and $S3$, 
a maximum occurs for $S4$ and $S5$.

\subsubsection{Trajectory of the capsule and membrane rotation} 

\begin{figure}
   \newskip\subfigcapskip \subfigcapskip=0pt
    \centering
    \hspace{-2.5em}{\label{fig:duct_xy_prof_v_ca}
    \includegraphics[scale = 1.15]
{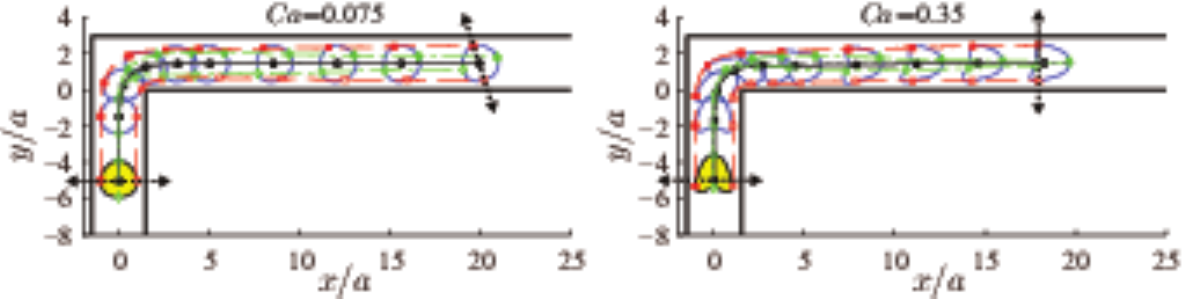}\hspace{0em}
 }
 \caption{(Colour online) Trajectories and profiles on the $x-y$ plane
   of capsules with capillary number  $Ca=0.075$ (left) and
   $Ca=0.35$ (right), reduced bending modulus $Cb=0.04$ and confinement
   $H_{x}/a=3$. The yellow shading denotes the initial equilibrium shape. 
   The black curve with circles represents
   the centroid path. The grey dash-dotted curve is the centreline streakline of
   the single-phase flow. Dash-dotted green curves with filled and hollow 
diamonds show trajectories of front and rear apices, respectively; dashed red
curves with squares stand for that of left and right apices. The dashed arrows connecting
the left and right apices indicate the rotation of the membrane.}
   \label{fig:duct_v_ca_profile}
 \end{figure}

The deformation and trajectories of capsules with $Ca=0.075$ (left) and 
$Ca=0.35$ (right) are displayed in figure~\ref{fig:duct_v_ca_profile}. 
The centroid trajectory (black curves with circles) closely
match the middle streakline $S3$ (dash-dotted grey curve) and is almost insensitive to the membrane elasticity. 
We also mark and trace the four apices of the capsule from the equilibrium shape. 
For $Ca=0.075$, we identify a clear rotation by comparing the initial and final positions of the apices.
The front and rear apices initially 
on $S3$, follow trajectories (indicated by filled and
hollow diamonds respectively) deviating from $S3$ significantly; the
front/rear apex drifts towards the outer/inner corner, eventually
remaining above/below the centroid trajectory. The left/right apex
starts from the same vertical position and approximately moves along
the streakline $S1$/$S5$. These are characterised by a
decreasing/increasing velocity around the corner (see
figure~\ref{fig:streamline_background}); as a result, the right apex travels
beyond the left, as shown in
figure~\ref{fig:duct_v_ca_profile}. 
The material points on the capsule rotate therefore in the anti-clockwise direction. This rotation is induced by the
flow near the corner: this is spatially nonuniform across the duct and the material points near the inner/outer 
corner are advected by the accelerating/decelerating flow, which result in
a net membrane rotation. 
In the case of $Ca=0.3$, the membrane rotation is not as clear. Compared to the case with $Ca=0.075$, the right apex 
is closer to the wall where the underlying flow is slower, thus compensating the 
increase of the fluid velocity near the corner. Hence, 
the left and right apices are roughly found at the same streamwise location downstream of the corner.

\subsection{Velocity of the capsule}\label{sec:velocity}
 \begin{figure}
    \centering
    \subfigure[]{\label{fig:duct_vel_ca_noscaled}
    \includegraphics[scale = 0.45]
{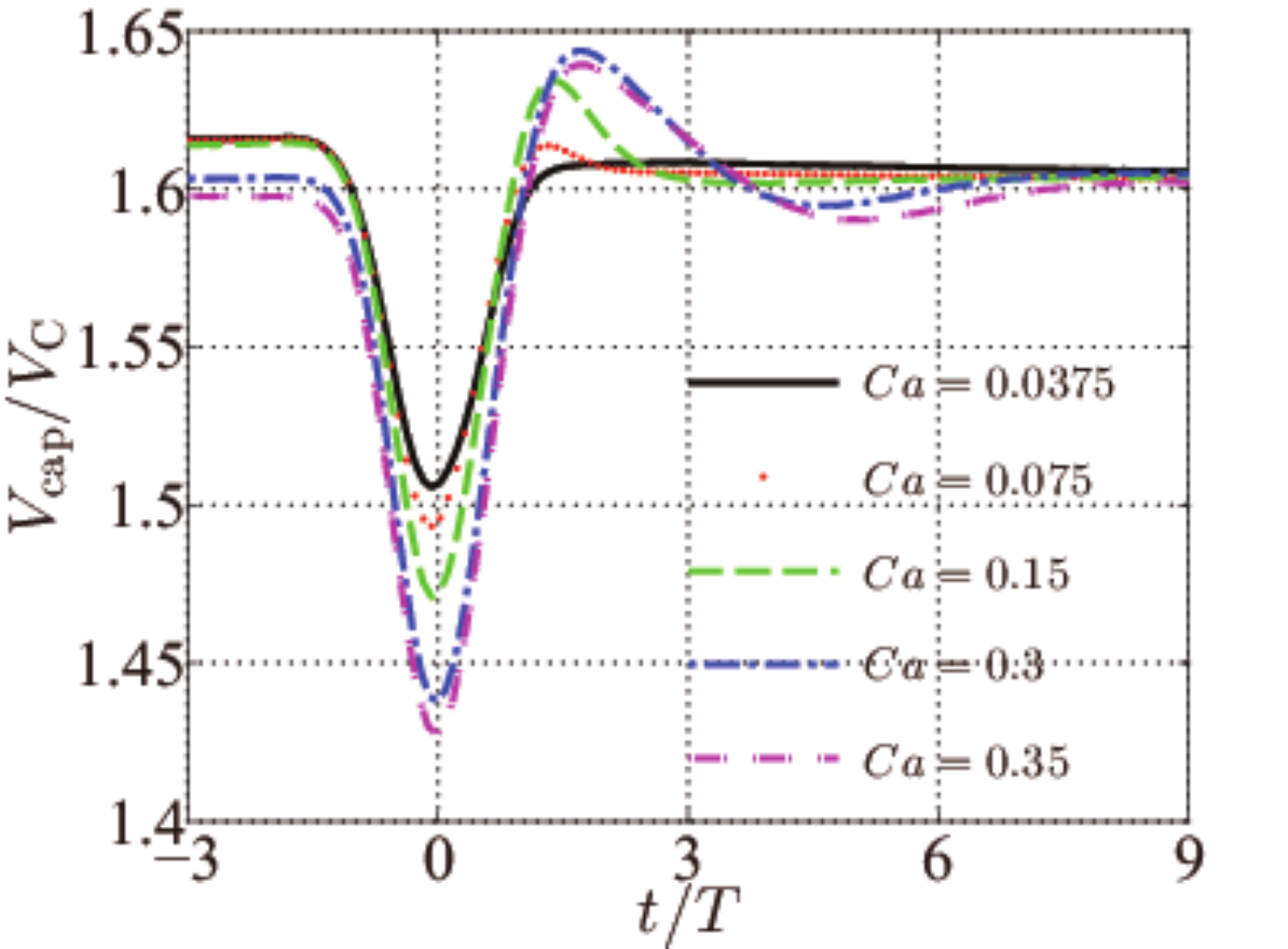}\hspace{0em}
 }
    \subfigure[]{\label{fig:duct_vel_ca_withback}
    \includegraphics[scale = 0.45]
{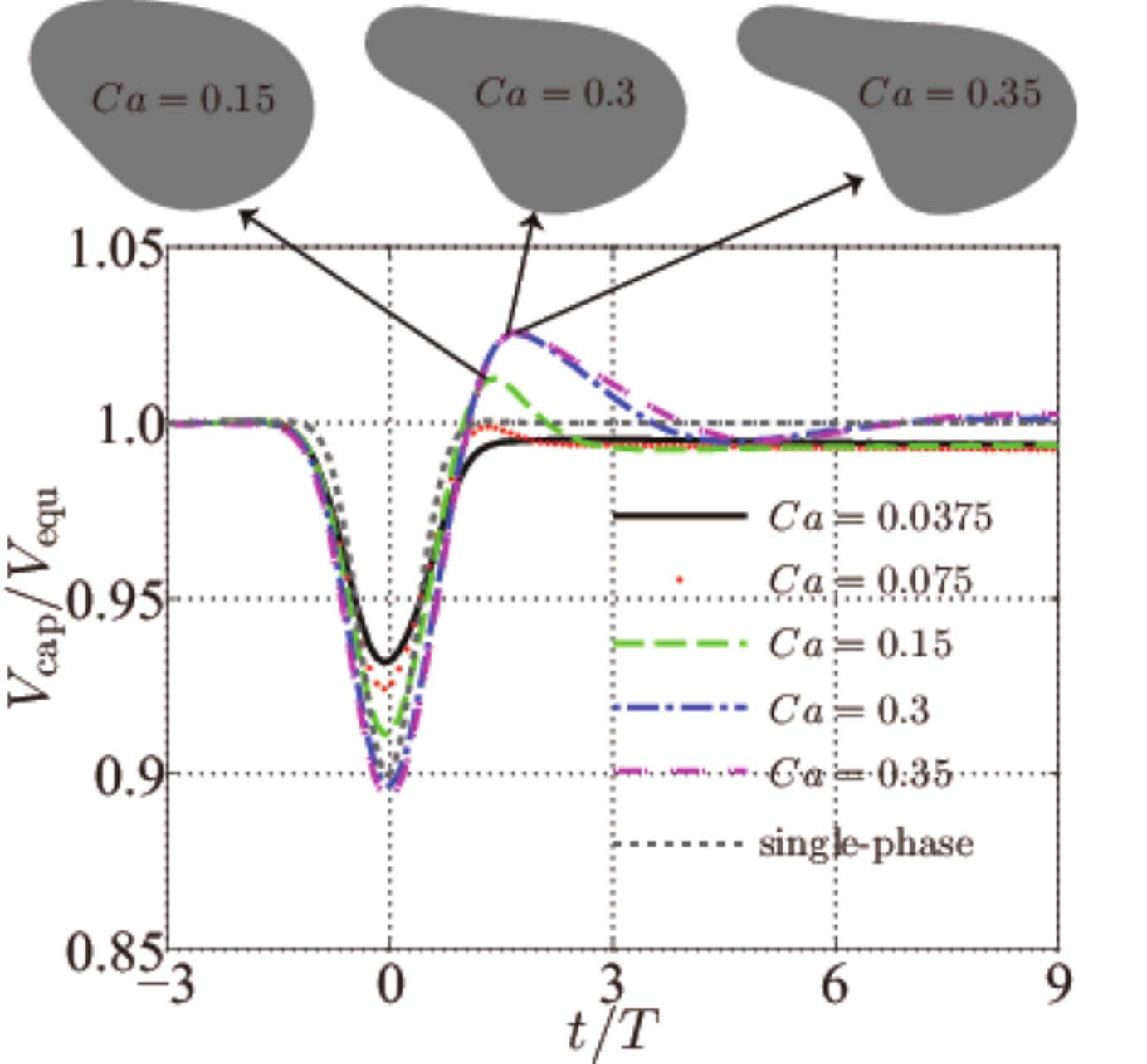}
 }
   \caption{(Colour online) Time evolution of the velocity of the
     capsule centre, $V_{\mathrm{cap}}$, scaled by the mean velocity $V_{\mathrm{C}}$ of the duct
     in~\subref{fig:duct_vel_ca_noscaled}, and by the cell velocity at equilibrium 
$V_{\mathrm{equ}}$
     in~\subref{fig:duct_vel_ca_withback}. The confinement $H_{x}/a=3$ and results are shown for capsules with 
$Ca=0.0375$,
     $0.075$, $0.15$, $0.3$ and $0.35$ and a reduced bending modulus $Cb=0.04$.
     The shape of the capsules at the 
maximum velocity is provided 
in~\subref{fig:duct_vel_ca_withback}.}
   \label{fig:duct_vel_ca}
 \end{figure}
 
The velocity of the capsule centre, $V_{\mathrm{cap}}$, scaled by the mean velocity $V_{\mathrm{C}}$ is reported in 
figure~\ref{fig:duct_vel_ca_noscaled} as a function of time; the same quantity instead divided by the equilibrium 
velocity $V_{\mathrm{equ}}$, 
is depicted  in figure~\ref{fig:duct_vel_ca_withback},
together with the velocity on the centre streakline $S3$ of the single-phase flow (cf.\ figure 
\ref{fig:streamline_background}). All
capsules move faster than the average flow velocity, a signature of the F\r{a}hraeus effect.
 Note that the equilibrium velocity $V_{\mathrm{equ}}$, the velocity in a straight duct,  decreases slightly with 
$Ca$.
  This was also discussed by \citet{dimitra11_squre} (fig.8a  in their paper):  $V_{\mathrm{equ}}$ increases with 
$Ca$ as $H_{x}/a=2.5$ but
 decreases in the less confined case, $H_{x}/a=\frac{10}{3}$; our simulations with $H_{x}/a=3$ are between the two cases 
in \citet{dimitra11_squre} and confirm the negative trend of $V_{\mathrm{equ}}$ at low confinement.
The velocity of the capsule is related to the thickness of the capsule-wall lubrication film; 
a thinner film induces higher viscous dissipation and thus reduces the capsule velocity. Indeed, the thickness of the 
film when the capsule is slower ($Ca=0.35$),
 is about $93\%$ that of the fast capsule ($Ca=0.0375$).

The velocity of the capsule 
decreases when approaching the corner and increases when leaving it, reflecting the behaviour of the background flow; 
the time histories reveal  a minimum located at $t=0$, when the particle centre is on the corner axis. This minimum 
velocity decreases 
 with the capillary number $Ca$; indeed, a slightly thinner lubrication film is observed at the corner axis
as $Ca$ changes from $0.35$ to $0.075$ (see fig.~\ref{fig:duct_v_ca_profile}). Unlike the underlying flow, the
motion of the capsule clearly breaks the time-reversal symmetry about $t=0$, revealing an overshoot during the 
recovery
stage; this symmetry breaking becomes more evident for higher $Ca$. This loss of symmetry is related to 
the viscoelasticity induced by the fluid-capsule interaction. 

We report the shape of the capsule associated to the larger velocity overshoots ($Ca=0.15$,
$0.3$ and $0.35$) in
figure~\ref{fig:duct_vel_ca_withback} at the time the peak velocity is attained. 
A clear tail-like 
protrusion is observed for the  two largest $Ca$s, due to the streamwise stretching induced by
the background accelerating flow. Such a shape  is responsible for the observed velocity overshoot, 
as the streamwise membrane extension corresponds to a decrease of the cross-flow extension of the capsule (its volume 
must 
be conserved). 
This causes a larger  distance between capsule and wall and a reduced viscous dissipation. Not surprisingly, 
as the capsule leaves the corner,  its vertical 
dimension recovers to the equilibrium value and so does the velocity.

The velocity does not converge exactly to its equilibrium value, a maximum relative difference of around 
$0.6\%$ is observed. It would require a prohibitively long computational domain and integration time to 
obtain a precise convergence as also observed by 
\citet{woolf11_branch}; the physics of the final capsule relaxation is therefore  
beyond the scope of the present investigation.

\subsection{Capsule surface area and deformation}\label{sec:area}
 \begin{figure}
    \centering
    \hspace{-3em}\subfigure[$A/4\pi a^{2}$]{\label{fig:duct_areaTT_ca}
    \includegraphics[scale = 0.5]
{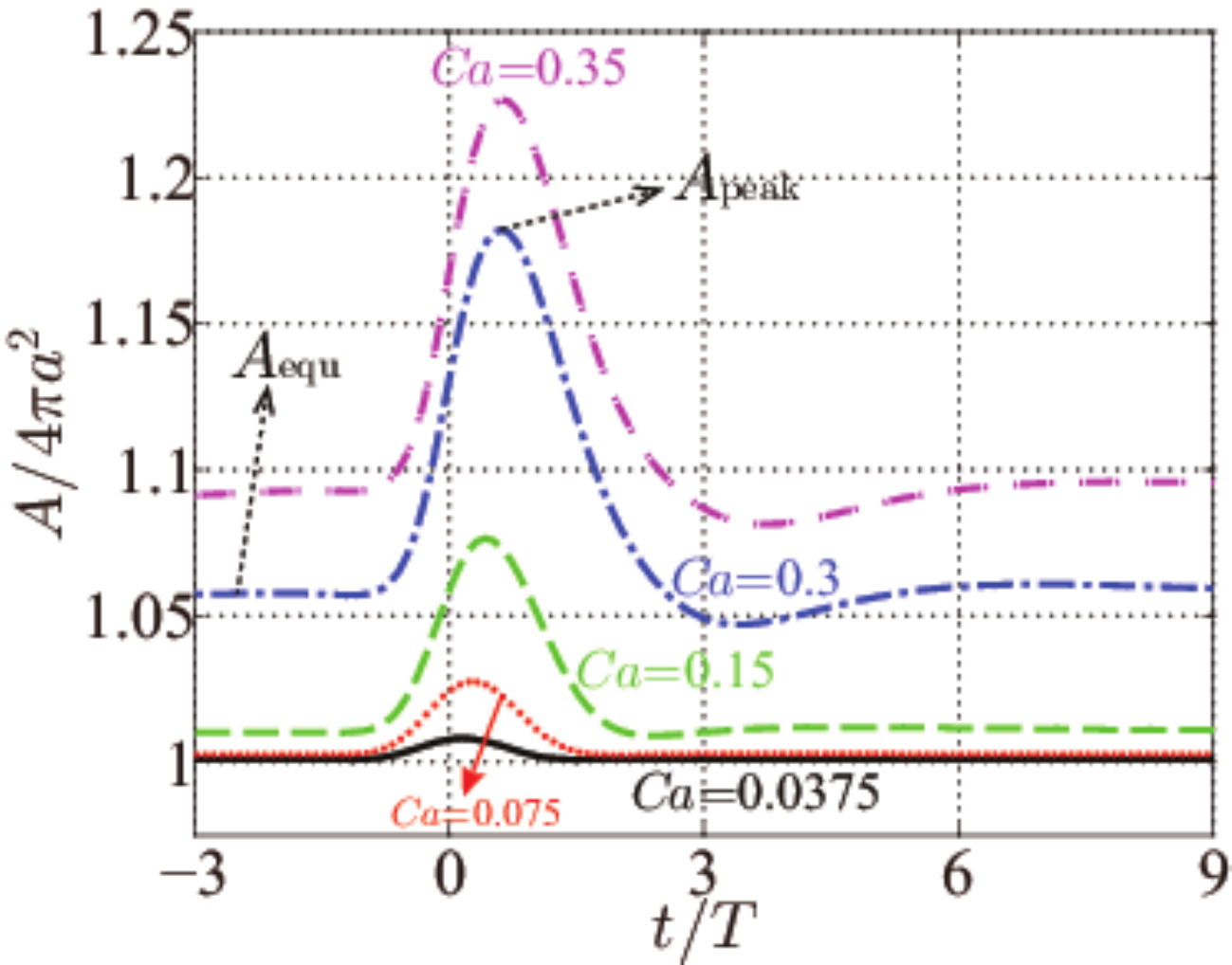}\hspace{0em}
 }
    \subfigure[$A_{xy}/\pi a^{2}$]{\label{fig:duct_areaXY_ca}
    \includegraphics[scale = 0.5]
{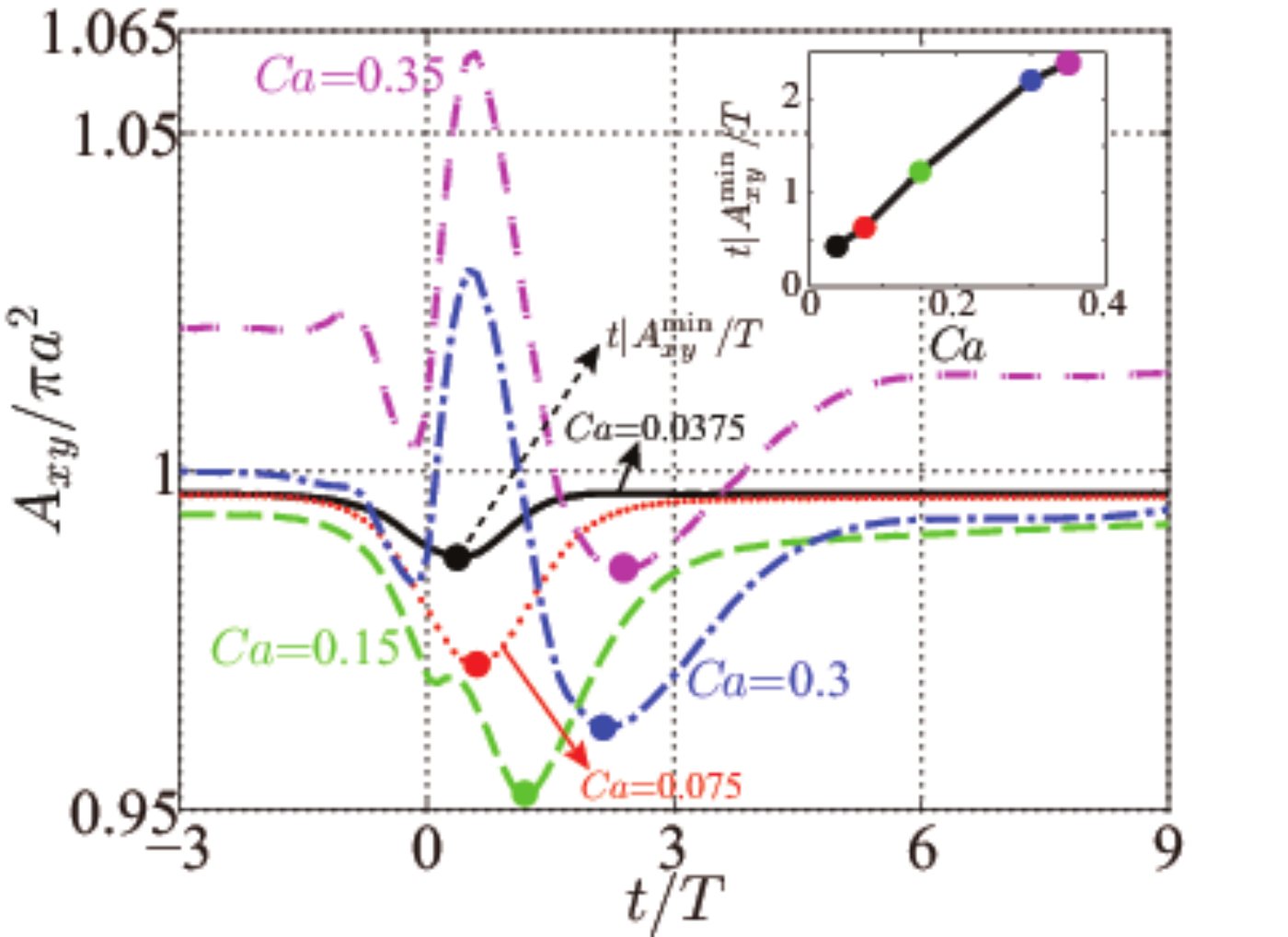}
 }
 \caption{(Colour online) Time evolution of the nondimensional surface area  for the same capsules in
 figure~\ref{fig:duct_vel_ca}. \subref{fig:duct_areaTT_ca}: total 
     surface area $A$ and \subref{fig:duct_areaXY_ca}:
     projected area on the $x-y$ plane, $A_{xy}$. $A_{\mathrm{equ}}$ and $A_{\mathrm{peak}}$ 
indicate the equilibrium and peak 
     value of $A$ respectively. Solid circles indicate the time 
$t|A_{xy}^{\mathrm{min}}/T$ when the minimum area $A_{xy}$ is achieved. The inset of 
\subref{fig:duct_areaXY_ca} shows $t|A_{xy}^{\mathrm{min}}/T$ versus $Ca$. }
   \label{fig:duct_area_ca}
 \end{figure}
 
The capsule surface area, $A$, is used as indicator of  the global deformation. 
This is reported in figure~\ref{fig:duct_areaTT_ca} for the same cases in figure~\ref{fig:duct_vel_ca}.
As the capsule is far away
from the corner, the area maintains the equilibrium value $A_{\mathrm{equ}}$, 
an increasing function of $Ca$. The area variation $A_{\mathrm{equ}}/4\pi a^{2}-1$ is almost zero
 for the cases with $Ca=0.0375$ and $0.075$, whereas 
it grows to values around $0.1$ when $Ca=0.35$. 
As the capsule travels around the corner, the deformation reaches its peak value
 and its variation $A_{\mathrm{peak}}/4\pi a^{2}-1$ is around $0.2$ for the highest $Ca$ investigated.

Ideally, the dependence of the area on the capillary number can be used to deduce the
membrane properties of capsules as shown by \citet{lefebvre2008flow,chu2011comparison,hu2013characterizing}, who
 focus on the identification based on deformation.
Nevertheless, direct measurement of the total surface area is not easy, while
it is more feasible to measure its two-dimensional  projection. As a consequence, 
we display the projection of the capsule area
 on the $x-y$ mid-plane in 
figure~\ref{fig:duct_areaXY_ca}.

The projected area $A_{xy}$ varies with time and cell deformability in
a more complicated way. 
For the two smallest values of $Ca$, $A_{xy}$ reaches the minimum around $t=0$ before
recovering to the equilibrium value past the corner. The cases characterized by   
$Ca=0.3$ and $0.35$ display a clear peak in deformation right after $t=0$, with two sharp troughs 
one before and one after. This wavy variation 
is already visible as $Ca=0.15$ although weak. 
Further examination of the behaviour in the range  $Ca \in [0.15,0.3]$ confirms 
that the time traces of the area deformation become 
more wavy as $Ca$ increases; indeed more elastic material 
is prone to exhibit more oscillatory motions under the same excitation, the spatially developing flow here. The inset 
of figure.~\ref{fig:duct_areaXY_ca} shows
the time $t|A_{xy}^{\mathrm{min}}/T$ corresponding to the minimum projected area $A_{xy}$. This can be 
regarded as the phase lag of the capsule and it increases almost linearly with $Ca$.

\subsection{Principal tension on the capsule} 
 The tension developing on the membrane is of great importance since it 
influences the release of molecules~\citep{goldsmith1995physical} and ATP~\citep{wan2008dynamics} by RBCs and 
causes
haemolysis, to cite two examples.  We analyse the principal tension $\tau_{i}^{P}
\left(i=1,2\right)$, to better understand the potential mechanical
damage of capsules passing through a corner. For any definition of strain energy function
$W_{S}\left(I_{1},I_{2}\right)$, $\tau_{i}^{P}$ are derived
as~\citep{skalak1973strain}:
\begin{eqnarray}
  \tau_{1}^{P} = 2\frac{\lambda_{1}}{\lambda_{2}}\left(\frac{\partial
W_{S}}{\partial I_{1}}+\lambda_{2}^{2}\frac{\partial W_{S}}{ \partial
I_{2}}\right),\nonumber\\
  \tau_{2}^{P} = 2\frac{\lambda_{2}}{\lambda_{1}}\left(\frac{\partial
W_{S}}{\partial I_{1}}+\lambda_{1}^{2}\frac{\partial W_{S}}{ \partial
I_{2}}\right).
\end{eqnarray}
We consider the major principal tension: 
$\displaystyle\max\left(\tau_{1}^{P}\left(\mathbf{x},t\right),\tau_{2}^{P}\left(\mathbf{x},t\right)\right)$ and the 
isotropic
principal tension $\left(\tau_{1}^{P}\left(\mathbf{x},t\right)+\tau_{2}^{P}\left(\mathbf{x},t\right)\right)/2$; their 
surface
maximum $\tau^{P}_{\mathrm{max}}\left(t\right)$ and
$\tpmaxiso\left(t\right)$ are defined as 
\begin{eqnarray}
\tau^{P}_{\mathrm{max}}\left(t\right)=\displaystyle
\max_{\substack{\mathbf{x},i=1,2}}\left(
\tau^{P}_{i}\left(\mathbf{x},t\right)\right),\\
\tpmaxiso \left(t\right)=\displaystyle
\max_{\substack{\mathbf{x}}}\left(\left(\tau^{P}_{1}\left(\mathbf{x},t\right)+
\tau^{P}_{2}\left(\mathbf{x},t\right)\right)/2 \right),
\end{eqnarray}
where $\left(t\right)$ will be omitted hereinafter for the sake of clarity. 

\begin{figure}
    \centering
    \includegraphics[scale = 0.275] {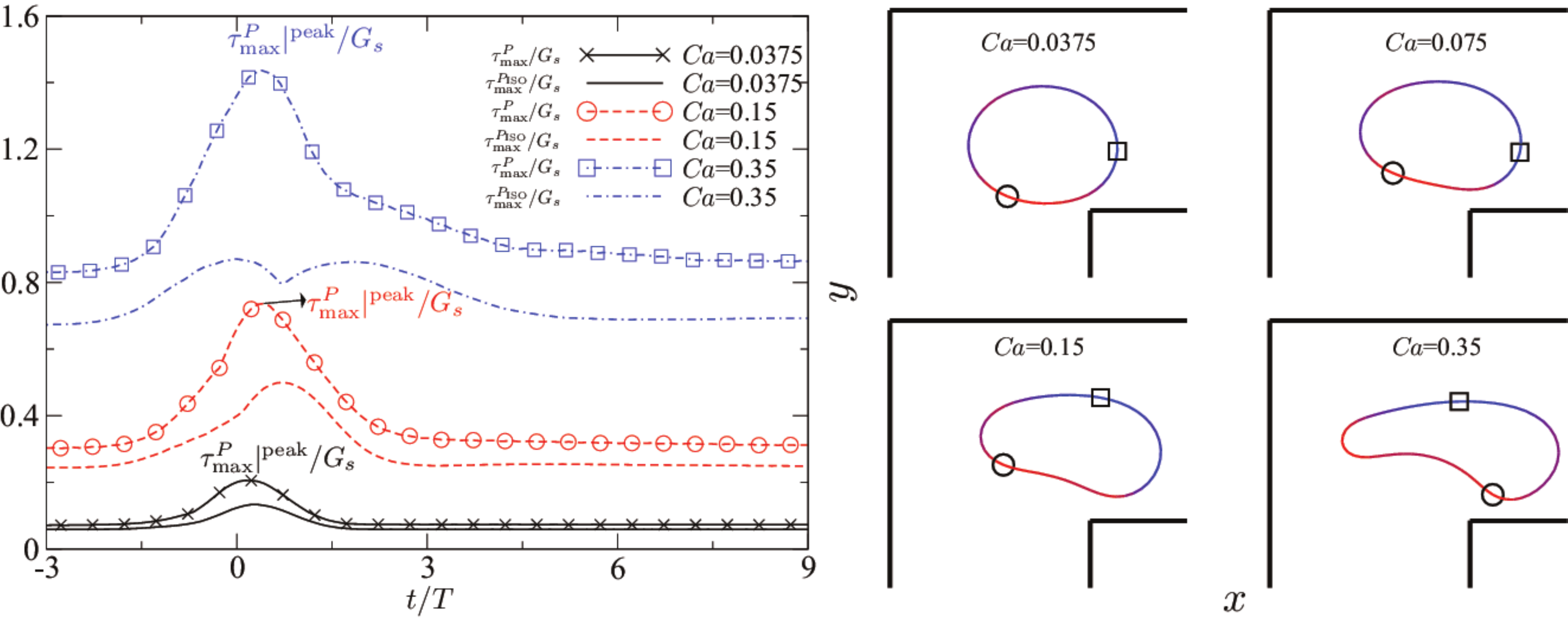}
   \caption{(Colour online) Left: time evolution of the maximum of the two principal tensions in
    the nondimensional form, $\tpmax/\gs$ and $\tpmaxiso/\gs$ for the major and isotropic principal tension 
respectively.
    Right: position and contour of the capsules on the $x-y$ plane when reaching the maximum major principal tension 
$\tpmax|^{\mathrm{peak}}$. The magnitude of $\tpmax$ is indicated by red/blue for low/high values and its 
minimum/maximum position by
the circle/square.}
   \label{fig:duct_prsts_ca}
\end{figure}

The temporal evolution of $\tpmax/\gs$ and $\tpmaxiso/\gs$ are shown in figure~\ref{fig:duct_prsts_ca} for  
capsules going through a straight corner. For most cases, both quantities increase monotonically with $Ca$, reaching
the peak values slightly after the corner before relaxing back to the equilibrium value. The 
difference between the two tensions is more pronounced at the corner, $\tpmax/\tpmaxiso \approx 2$, and
weak in the straight ducts. We also report in the figure the shape of some capsules when the maximum major principal 
tension is reached,
with the minimum and maximum of $\tpmax$ indicated by circles and square respectively. The maximum of $\tpmax$ develops 
in the front for the capsules as $Ca=0.0375, 0.075$, while it moves to the top part as $Ca=0.35$.
Material points are prone to accumulate in the rear of the capsule and the principal tension 
is therefore relatively low. As $Ca$ increases, the rear part of the capsule changes from a convex to concave shape, 
something more evident 
for the case $Ca=0.35$.

 \begin{figure}
    \centering
    \hspace{-3em}\subfigure[]{\label{fig:varywid_vel}
    \includegraphics[scale = 0.475]
{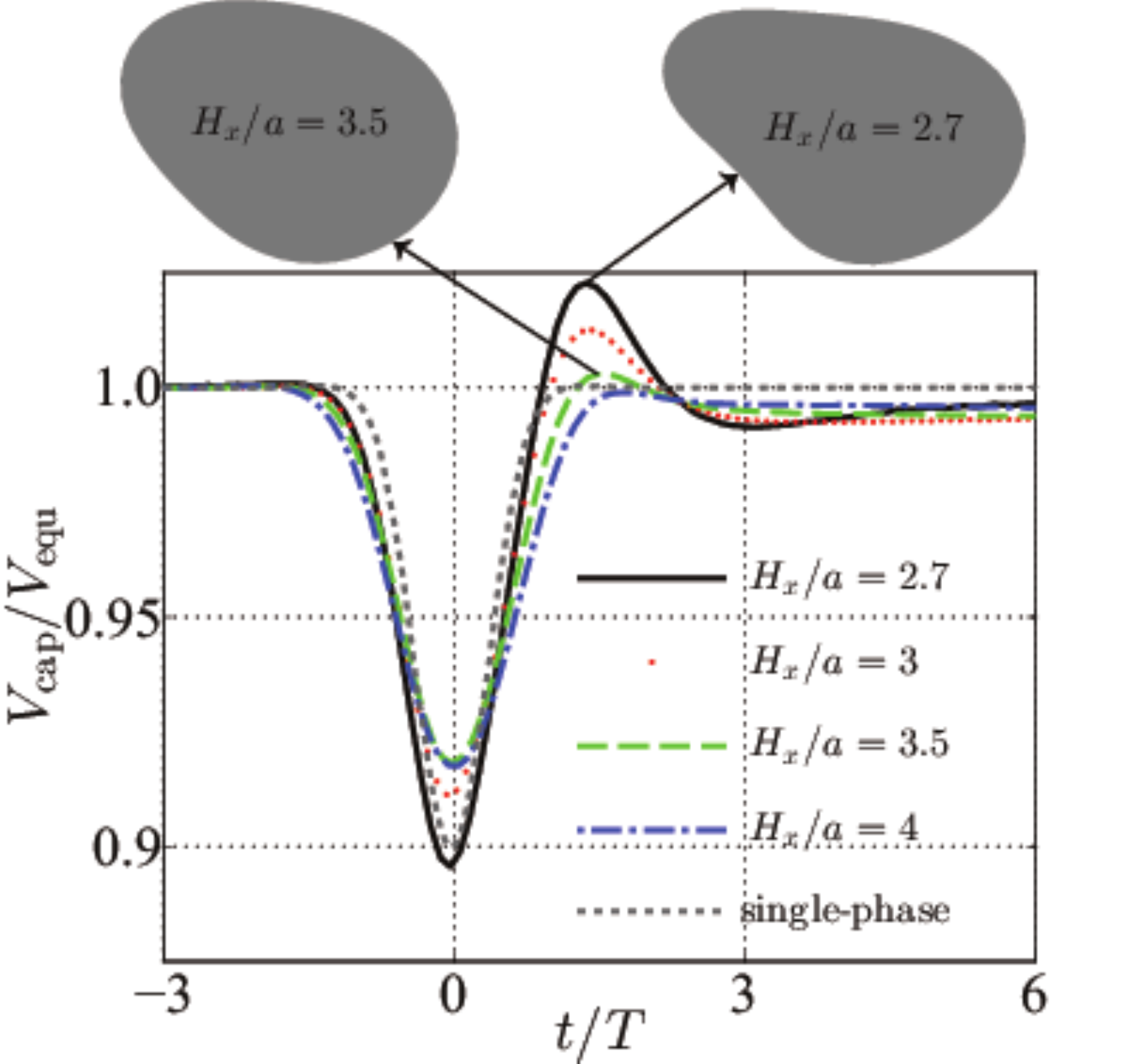}\hspace{0em}
 }
    \subfigure[]{\label{fig:varywid_prsts}
    \includegraphics[scale = 0.265]
{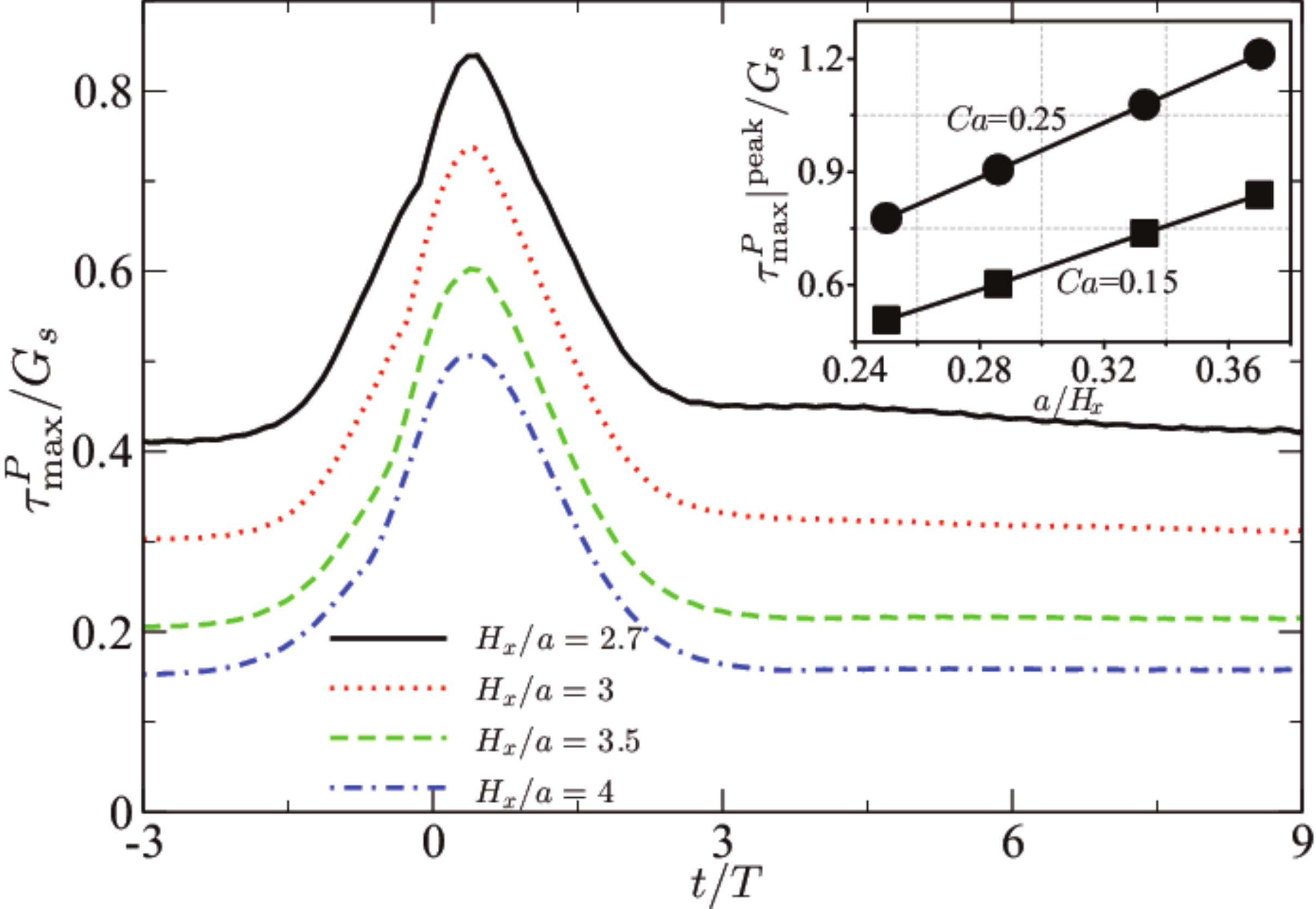}
 }
   \caption{(Colour online) Time evolution of \subref{fig:varywid_vel}: the velocity 
$V_{\mathrm{cap}}/V_{\mathrm{equ}}$  and \subref{fig:varywid_prsts}: the major principal 
tension $\tpmax/\gs$  for $Ca=0.15$ where the width of the square duct is varied from
$H_{x}/a=2.7$ to $H_{x}/a=4$. The shape of the capsule at the time when the maximum velocity is attained is also 
given in~\subref{fig:varywid_vel}; the inset of \subref{fig:varywid_prsts} shows the relation between the maximum major 
principal tension 
$\tpmax|^{\mathrm{peak}}/\gs$ and  the inverse of the duct width $a/H_{x}$ for $Ca=0.15$ and $0.25$.}
   \label{fig:varywid_vel_prsts}
 \end{figure}

\subsection{The influence of confinement and geometry of the corner}

We examine first the influence of confinement on the capsule motion by varying the
width $H_{x}/a$ from $2.7$ to $4$. The velocity
of the capsule, divided by its equilibrium velocity, is shown in figure~\ref{fig:varywid_vel} at $Ca =0.15$. The 
time-symmetry around $t=0$ is almost preserved for the least confined case $H_{x}/a=4$.
As the confinement increases, the symmetry breaking discussed before and the corresponding velocity overshoot become
more apparent. These are associated to a decrease of 
the minimum  velocity at the corner. As $H_{x}/a$ varies from $3.5$ to $2.7$, the velocity overshoot also clearly  
increases.
The shape of the capsule at the time of maximum velocity is also displayed in the figure. The capsule with
highest velocity is more elongated and has a larger  distance from the wall (lower lubrication friction), in analogy to 
the observations in section~\ref{sec:velocity} for capsules of different elasticity.
 
 The surface maximum of the nondimensional major principal tension $\tpmax/\gs$ is depicted in
 figure~\ref{fig:varywid_prsts} for the same cases: $\tpmax$ increases monotonically with 
the 
confinement. The maximum over time of $\tpmax|^{\mathrm{peak}}/\gs$  is displayed versus $a/H_{x}$ for
$Ca=0.15$ and $0.25$ in the inset of the same figure to show that
 the peak principal tension increases linearly with $a/H_{x}$. This relationship may be useful to estimate
 the mechanical stress/damage on the cells in micro-fluidic devices
already during the design stage.

  \begin{figure}
    \centering
    \includegraphics[scale = 0.3] {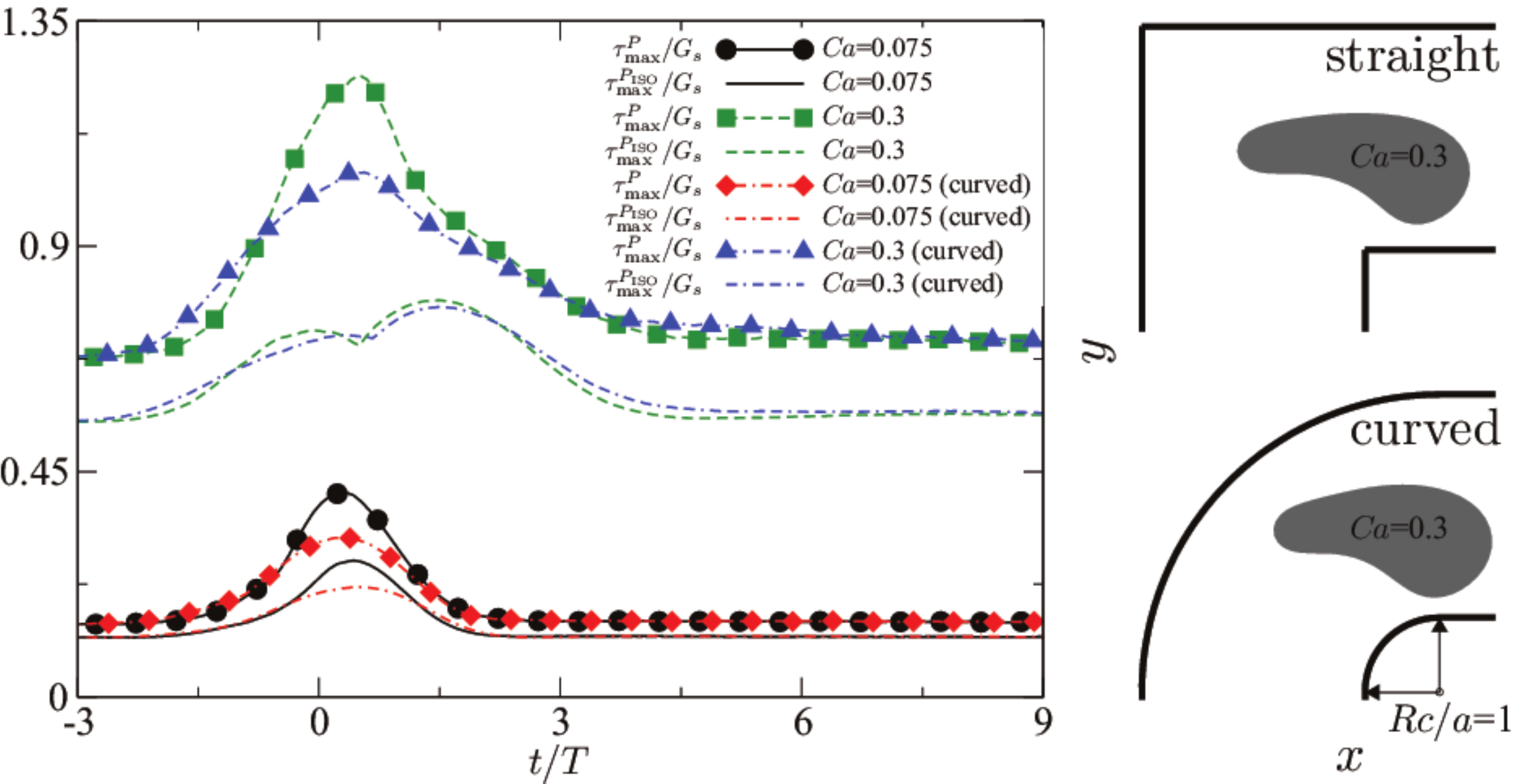}
   \caption{(Colour online) Time evolution of the major principal tension $\tpmax/\gs$ for
   capsules $Ca=0.075/0.3$  through a straight and a curved corner. The curvature radius
   of the curved corner is $R_{c}/a=1$. The shape of capsule on the $x-y$ plane is shown 
   when it reaches the peak $\tpmax$.}
   \label{fig:ductWithbent_prsts_ca}
 \end{figure}

Finally, we consider a curved corner with inner radius $R_{c}/a=1$.
 In figure~\ref{fig:ductWithbent_prsts_ca} we compare the principal tensions on the membrane with those in
the straight corner for capillary numbers $Ca=0.075$ and $0.3$. 
Except for the isotropic principal tension of the capsule $Ca=0.075$,  the principal tension decreases 
significantly in the curved corner.

\section{Discussion and Conclusion}\label{conclusion}
We investigate the motion of a three-dimensional deformable  capsule, whose membrane obeys the neo-Hookean 
constitutive relation, in a square duct flow with a corner. 
We present in this work a new implementation of the boundary integral method 
accelerated by the GGEM, the general geometry Ewald method, to resolve fluid-structure interactions at low Reynolds number in complex geometries.
The algorithm is coupled with a spectral method based on spherical harmonics for the membrane dynamics. 
In this section, we first discuss the details of the numerical method,
followed by a short summary of the main physical findings.

The GGEM shares similarities with the classic immersed boundary methods 
(IBM)~\citep{mittal2005immersed}. Both approaches require a Lagrangian mesh for the suspended objects and an Eulerian (typically 
Cartesian) mesh for the fluid; the Dirac delta function, representing the localized forcing from the object, is approximated numerically. 
In the IBM, the localized force is spread from each Lagrangian point onto a number of surrounding Eulerian
points to enforce the desired boundary conditions at the fluid/solid interface. 
The accuracy of IBM degenerates if close
hydrodynamic interactions arise, which may requires ad hoc corrections to account for the correct lubrication forces \citep{Iman-Picano}. As shown in Eq.~(\ref{eq:global_forcing}), the Dirac delta function is also smeared in 
the local problem of GGEM, but its singular behaviour can be solved accurately by boundary integral
techniques with singular integration. If a regularized-Stokeslet  technique is instead employed
as in \citet{gram_pof_pair_polymer,graham07_prl}, the GGEM closely resembles a IBM as proved 
by~\citet{gram_pof_pair_polymer}. Note also that traditional IBM requires a uniform Eulerian grid to conserve the 
moments of the force and sophisticated treatments are needed to adapt IBM to a non-uniform and/or unstructured 
grid as done by \citet{pinelli2010immersed} and \cite{mendez2014unstructured} among others. One advantage of the GGEM is that the
smoothing of the local forcing is exactly compensated by the global forcing due to the linearity of Stokes
equations. The integral of the force field and its moments are therefore preserved. Hence, 
Stokes solvers based on uniform or non-uniform grids can be readily coupled to the GGEM. In our case, the Eulerian grid points 
(GLL points) are non-uniformly distributed as shown in figure~\ref{fig:mesh_fd_cl}.

GGEM is originally designed to resolve the hydrodynamic interaction among multiple
particles in Stokes flows. Suppose to have $\np$ particles and each of them is discretized into $M$
Lagrangian points, the total number of points is $\np M$ and the number of degrees of freedom $\N \sim \np M$.
For traditional non-accelerated BIM, the number of operations required to form the mobility matrix
scales with $\N^{2}$. This poses the major difficulty in applying BIM to a large number of
particles. Accelerating techniques for BIM have thus been developed to overcome this restriction, and GGEM is one of 
them. The decomposition of the Dirac delta function into two parts reduces the 
number of operations from $O\lp \N^{2} \rp$ to $O\lp \N \rp$ or $O\lp \N \log{\N}\rp$~\citep{graham07_prl}.
The modified Green's function for the local problem is designed such that the local solution decays exponentially over 
a distance of about $\alc^{-1}$. Neglecting interactions that occur beyond the cut-off distance $\rcut \sim \alc^{-1}$, 
the number of operations for the local solution decreases and scales linearly with $\N$. The scaling of the global problem depends 
on the mesh-based solver, as well as the geometry and boundary conditions of the computational domain. The solver used 
here, NEK5000, is computationally more expensive than Fourier-based methods as those used in \citet{graham12_jcp}, but 
it allows for arbitrary geometries and is highly parallel.

It is hard  to provide a scaling for the global part of the problem in general geometries, however the advantage of a numerical approach like that pursued here relies on two points: i) GGEM provides a convenient way to reshape the $O\lp \N^{2} \rp$ long-ranged
interactions and pack them into a problem solvable by a mesh-based solver; ii) a highly parallel solver is chosen
to considerably reduce the computational time. Note that a naive parallelization of traditional BIM implementations is not
possible for a large $\N$ due to the prohibitively large amount of memory needed and poor scalability of the
linear system with a dense matrix. It should be said that our implementation
might be less efficient than traditional BIM to study the dynamics of one (as we do here) or a few capsules. 
This is however our first step in the development of a computational framework for suspensions of deformable/rigid
particles in general geometries.

This numerical approach is used here to examine the motion of a capsule through a 
square duct with a corner, focusing on its trajectory, velocity, deformation, total and projected surface area, and 
principal tension. We aim to better understand the transient 
dynamics of capsules in a micro-fluidic device with realistic geometries.

We study the deformation of the capsules when varying the capillary number, the ratio of viscous to elastic forces. 
The capsule trajectories closely follow 
the underlying flow and are therefore rather insensitive to how the capsules deform. Due to the strong confinement, 
deviations from the 
underlying streamline requires a significantly viscous dissipation. Conversely, the deformability of a capsule closely influences 
its shape, velocity and the mechanical stress developing on the membrane as documented in the results section.

The corner flow can  be potentially adopted to infer the material properties of deformable particles 
as shown by \citet{lefebvre2008flow} and \cite{chu2011comparison} using straight tube or channel flows. Unlike these
 works, transient effects are present in the flow past a corner because of its spatial
inhomogeneity. When the capsule is far away from the corner, 
the surface area, velocity and principal tension reach equilibrium values that are function of the capillary number $Ca$. 
When flowing around the corner, the membrane area and tension reach their maxima 
while the velocity the minimum; these extrema are shown here to clearly depend on $Ca$. 
By utilizing a spatially developing flow, 
the shape and/or velocity of the capsules can be measured not only at the equilibrium state but also during the
transient motions. More robust and accurate inverse methods may be developed using measurements of the
extrema values. We further note a new time scale is introduced in the corner flow, hence the phase 
lag of the capsule can be identified as illustrated by the temporal evolution of the projected area (see 
figure.~\ref{fig:duct_areaXY_ca}); this quantity indicating the viscoelasticity of capsule is not accessible from the 
traditional steady flow experiments. The spatially developing flow indeed provides supplemental information 
characterising the material properties of natural and synthetic cellular structures.

The capsule shape is also closely linked to its velocity. For low 
$Ca$,  the velocity is similar  to that of the underlying flow, with an almost perfect time-reversal symmetry;
as $Ca$ increases, i.e.\ more pronounced deformations, this symmetry is broken and a velocity overshoot appears past the corner. 
The streamwise elongation of the capsule increases the capsule-wall distance, and the corresponding lower viscous 
dissipation can explain the  higher capsule velocity. 

The surface maxima of the major and isotropic principal tension become significantly different only when the 
capsule is flowing around the corner. During this time, the maximum major principal tension appears in the front for  
capsules for configurations with low $Ca$, and shifts towards the outer edge as $Ca$ increases.

We have also examined the influence of confinement and of the geometry of corner. We identify a positive correlation
 between the asymmetry of the velocity profile and the level of confinement. The peak of the
major principal tension increases linearly with the inverse of the duct width $a/H_{x}$. Finally, we show that a curved corner 
reduces the  major principal tension and the deformation of the capsule. We believe the present work can improve our understanding of the capsule motion in complex geometries and support the
the design of micro-fluidic devices with multiple corners and branches.

\section*{Acknowledgements}
We thank Prof. Dominique Barth{\`e}s-Biesel, Prof. Michael D. Graham, Prof. Jonathan B. Freund and Dr. Hong Zhao for 
useful 
discussions. Funding by
VR (the Swedish Research Council), the Linn\'e FLOW Centre at KTH and computer time via SNIC (Swedish National 
Infrastructure for Computing) and HPC at EPFL are greatly acknowledged. Lailai Zhu acknowledges the financial support 
from the European Research Council (ERC) grant  'simcomics-280117' as a postdoc researcher at EPFL where part of the 
work is performed. This research is also supported by the ERC Grant 
'2013-CoG-616186, TRITOS' to Luca Brandt.

\bibliographystyle{jfm}

\end{document}